\documentclass[preprint]{revtex4}
\usepackage{graphicx}
\def\ben{\begin{enumerate}}
\def\een{\end{enumerate}}
\def\ie{{\it i.e.}}
\def\ra{\rightarrow}
\def\to{\ra}

\def\be{\begin{equation}}
\def\ee{\end{equation}}
\def\bea{\begin{eqnarray}}
\def\eea{\end{eqnarray}}
\def\tbtb{t\anti b \, \anti t b}
\def\bbbb{b\anti b b\anti b}
\def\wt{\widetilde}
\def\epem{e^+e^-}
\def\bb{b\anti{b}}

\def\bbA{\bb \ha}
\def\tanb{\tan\beta}
\def\cotb{\cot\beta}

\def\eps{\epsilon}
\def\hl{h}
\def\hh{H}
\def\ha{A}
\def\hp{H^+}
\def\hm{H^-}
\def\hpm{H^\pm}
\def\mhpm{m_{\hpm}}
\def\anti{\overline}
\def\mhh{m_{\hh}}
\def\mha{m_{\ha}}
\def\mhl{m_{\hl}}
\def\mz{m_Z}
\def\gev{~{\rm GeV}}
\def\tev{~{\rm TeV}}
\def\fbi{~{\rm fb}^{-1}}
\def\fb{~{\rm fb}}
\def\call{{\cal L}}
\def\rts{\sqrt s}

\def\vev#1{\langle #1 \rangle}
\def\br{{\rm BR}}
\def\gamhatot{\Gamma_{\rm tot}^{\ha}}
\def\gamhhtot{\Gamma_{\rm tot}^{\hh}}
\def\gamhpmtot{\Gamma_{\rm tot}^{\hpm}}
\def\gamres{\Gamma_{\rm res}}
\def\mt{m_t}
\def\mb{m_b}
\def\lsim{\mathrel{\raise.3ex\hbox{$<$\kern-.75em\lower1ex\hbox{$\sim$}}}}
\def\gsim{\mathrel{\raise.3ex\hbox{$>$\kern-.75em\lower1ex\hbox{$\sim$}}}}
\def\ifmath#1{\relax\ifmmode #1\else $#1$\fi}
\def\half{\ifmath{{\textstyle{1 \over 2}}}}

\def\beq{\begin{equation}}
\def\eeq{\end{equation}}
\def\bit{\begin{itemize}}
\def\eit{\end{itemize}}
\begin{document}

\bibliographystyle{revtex}

\preprint{
 {\vbox{
% \hbox{CERN--TH/00-}
 \hbox{UCD--02--16}
 \hbox{MADPH--02--1307}
 \hbox{ANL-HEP-PR-02-090}
 \hbox{hep-ph/0212151}}}}

\vspace*{2cm}

\title{Determining \boldmath$\tanb$ with Neutral and Charged Higgs Bosons\\ 
at a Future \boldmath$e^+e^-$ Linear Collider}

\author{J. Gunion$^1$, T. Han$^2$, J. Jiang$^3$, A. Sopczak$^4$}
\affiliation{\vspace*{0.1in}
$^1$Davis Institute for HEP, Univ.~of California, Davis, CA 95616 \\
$^2$Dept.~of Physics, Univ.~of Wisconsin, Madison, WI 53706\\
$^3$HEP Division, Argonne National Laboratory, Argonne, IL 60439\\
$^4$Lancaster University, UK
\vspace*{1cm}}


\begin{abstract}
\vspace*{1cm}
The ratio of neutral Higgs field vacuum expectation values, $\tanb$,
is one of the most important parameters to determine
in either the Minimal Supersymmetric Standard Model (MSSM)
or a general type-II Two-Higgs Doublet Model (2HDM).
Assuming an energy and integrated
luminosity of $\rts=500\gev$ and $\call=2000\fbi$ at a future linear
collider (LC), we show that a very accurate determination of $\tanb$ 
will be possible for low and high $\tanb$
values by measuring the  production rates of Higgs bosons
and reconstructing Higgs boson decays. In particular,
based on a TESLA simulation, and
assuming no other light Higgs bosons and $100\leq \mha\leq 200\gev$, 
we find that the rate for the process $\epem\to\bb\ha\to\bb\bb$ 
provides a good determination
of $\tanb$ at high $\tanb$. In the MSSM
Higgs sector, in the sample case of $\mha=200\gev$,
we find that the rates for $\epem\to\bb\ha+\bb\hh\to \bb\bb$
and for $\epem\to \hh\ha\to\bb\bb$ provide a good determination
of $\tanb$ at high and low $\tanb$, respectively. 
We also show that the direct measurement of the average total widths
of the $\hh$ and $\ha$ in  $\epem\to\hh\ha\to \bb\bb$ events provides
an excellent determination of $\tanb$ at large values. 
In addition, the charged Higgs boson process
$\epem\to \hp\hm\to t\anti b \anti t b$ has been studied.
The sensitivity to $\tanb$ at the LHC obtained directly from
heavy Higgs boson production is briefly compared to the LC results.
\end{abstract}

\maketitle

%\vspace*{-1.3cm}
\vspace*{.1cm}
\section{Introduction}

\vspace*{-4mm}
Theories beyond the Standard Model (SM)
that resolve the hierarchy and fine-tuning problems
typically involve extensions of its
single-doublet Higgs sector to at least a 
two-doublet Higgs sector (2HDM) \cite{Gunion:1989we}. 
The most attractive such model is the Minimal Supersymmetric 
Standard Model (MSSM), which contains a constrained two-Higgs-doublet 
sector \cite{Gunion:susyh}. In other cases, the effective
theory below some energy scale is equivalent to a 2HDM extension
of the SM with no other new physics. Searching for the Higgs particles
and studying their properties have high priority for both
theoretical and experimental activities in high energy physics.

Among other new parameters in 2HDM and SUSY theories, 
one is of particular 
importance: the ratio of the vacuum expectation values
of the two Higgs fields, commonly denoted as 
$\tanb=v_2/v_1$. It characterizes the relative fraction 
that the two Higgs doublets contribute to the electroweak 
symmetry breaking $v^2=v^2_1+v^2_2$, where $v\approx 246$ GeV.
The five physical Higgs states couple to the fermions 
at tree-level \cite{Gunion:1989we,Gunion:susyh} as
\vspace*{-2mm}
\begin{eqnarray}
&&\hl\bar tt :  -i\frac{m_t}{v}\frac{\cos\alpha}{\sin\beta} \approx
       -i\frac{m_t}{v}\quad \qquad \qquad
\hl\bar bb :  i\frac{m_b}{v}\frac{\sin\alpha}{\cos\beta} \approx
       -i\frac{m_b}{v}\label{equ:hff}\\ 
&&\hh\bar tt :  -i\frac{m_t}{v}\frac{\sin\alpha}{\sin\beta} \approx
       i\frac{m_t}{v}{\cot\beta}\quad \qquad
\hh\bar bb :  -i\frac{m_b}{v}\frac{\cos\alpha}{\cos\beta} \approx
       -i\frac{m_b}{v}{\tan\beta}\label{equ:Hff}\\ 
&&\ha\bar tt :  
-\frac{m_t}{v}\cot\beta\ {\gamma}_5 \qquad \qquad\qquad\qquad\ 
\ha\bar bb :  -\frac{m_b}{v}\tan\beta\ {\gamma}_5 \label{equ:Aff}\\ 
&&H^+\bar{t}b :  i\frac{V_{td}}{\sqrt{2}v}[ 
m_b\tanb (1+\gamma_5) + m_t\cotb (1-\gamma_5) ],
\label{equ:Htb}
\end{eqnarray}
where $\alpha$ is the mixing angle in the CP-even sector,
and the approximation indicates the decoupling limit for
$\mha\gg \mz$ in the MSSM \cite{howie,Gunion:2002zf}, in which the 
couplings of the light Higgs boson $h$ become SM-like. 
Eqs.~(\ref{equ:Hff})--(\ref{equ:Htb}) show that $\tanb$ 
governs the coupling strength of
Yukawa interactions between the fermions and the heavy Higgs bosons.
In fact, heavy Higgs boson measurements sensitive
to their Yukawa couplings are far and away the most {\it direct} 
way to probe the structure of the vacuum state of the model
as characterized by the ratio of vacuum expectation values
that defines $\tanb$.

The parameter $\tanb$ enters all other sectors of the
theory in a less direct way \cite{Gunion:1989we}. 
For instance, in supersymmetric 
theories the interactions of the SUSY particles have $\tanb$ dependence. 
In addition, the relations of SUSY particle masses to the soft 
SUSY breaking parameters of supersymmetry involve $\tan\beta$.
The renormalization group evolution of the Yukawa 
couplings from the unification scale to the electroweak scale 
is sensitive to the value of $\tan\beta$. The large top  
quark mass can be naturally explained with $m_b-m_\tau$ unification 
as a quasi-infrared fixed point of the top Yukawa coupling if 
$\tan\beta\simeq 2$ or $\tan\beta \simeq 56$ \cite{BBO}.  
The possibility of SO(10) Yukawa unification  
requires high $\tan\beta$ solutions \cite{so(10)}. 
The predicted mass of the lightest SUSY Higgs boson also 
depends on $\tan\beta$, with a higher mass at larger
$\tan\beta$ \cite{deltamh}. 
It will be very important to compare the
measurements of and constraints on $\tan\beta$ 
from these other sectors of the theory
to the direct determination of $\tanb$ coming from
the heavy Higgs boson measurements that depend fundamentally
on $\tanb$ through the Yukawa couplings.

Currently, some regions of the MSSM parameter space have been excluded 
at LEP due to the lower bound on the lightest Higgs boson mass.
(A review of LEP-1 Higgs results shows possible signatures
for all neutral and charged Higgs boson search channels~\cite{lep1}.)
Particularly interesting is the exclusion $0.5<\tan\beta<2.4$
in the maximal-top-squark-mixing scenario when the SUSY
scale $M_{\rm SUSY}\lsim 1\tev$ \cite{lepwg}. 
More general MSSM parameter scans reduce the excluded $\tanb$ 
range~\cite{generalscan}, especially if the top quark mass
is allowed to vary within its $1\sigma$ error range.
Searches for top decay $t\to H^+b$ at the Fermilab Tevatron, 
sensitive to $\tanb>50$ with $m_t>\mhpm$, and searches for the final
state $b\bar b h \to b\bar b b\bar b$, sensitive
to $\tanb>35$, have also set limits \cite{tev}
on the very high values of $\tanb$ as a function of the Higgs mass. 
Precision electroweak measurements may provide some weak
constraints on $\tan\beta$ \cite{yamada}.
Much of the parameter space, however, remains to be explored at future
collider experiments. 

Because of the significance of $\tan\beta$ for the theory and 
phenomenology, it is important to constrain it and eventually
to determine its value in future collider experiments.
At the upgraded Tevatron (Run II) with high
luminosity (10 $\fbi$ or higher), complementarity among the processes
$q\bar q \to Wh,Zh$ \cite{mmw}, 
$gg \to b\bar b h,b\bar b H,b\bar b A$ \cite{hty,run2}, 
and $gg\to h,H,A\to \tau^+\tau^-$ \cite{bhr} may allow 
SUSY Higgs detection throughout the full 
SUSY parameter $(\mha,\tanb)$ plane. 
Depending upon the integrated luminosity and the value of $\tanb$, either
we will be able to directly observe the heavy Higgs processes
and be able to determine $\tanb$ or, if the $\hh,\ha,\hpm$ are not
detected, we will be able to place an upper bound on $\tanb$ as
a function of $\mha$. At high $\tanb$, SUSY particle production may
provide an additional handle \cite{susytev} for exploration of the 
parameters as well. More recently, it has been pointed
out that a large value of $\tanb$ can substantially enhance  
$B$ meson decay branching fractions \cite{chris} and thus could
enhance our ability to probe SUSY parameters
through indirect SUSY and Higgs signals \cite{uli}.
SUSY Higgs detection at the LHC has been studied for many years;
see \cite{Gunion:1996cn} for a summary of the early work. Recent
studies by the LHC CMS and ATLAS collaborations can be found in
\cite{LHCcms,LHCatlas,dieter}. The conclusion of these studies
is that there is a ``no-lose'' theorem for SUSY Higgs discovery
at the LHC, although there is a substantial region of 
$(\mha,\tanb)$ parameter space where only the light CP-even $\hl$
will be detected. The determination of $\tanb$
at the LHC by measuring Higgs boson production rates was analyzed
in \cite{Gunion:1996cn} and has been explored in greater detail
in \cite{LHCatlas,Assamagan:2002ne}. The possibility of
measuring $\tanb$ at the LHC via production of gauginos and squarks 
was explored in \cite{baer,lhctanb}. 
A future linear collider has great potential for discovering
new particles and measuring their properties due to its clean
experimental environment. An early study \cite{nlctanb} has been 
made to determine $\tanb$ via gaugino production in $e^+e^-$
collisions, followed by a discussion for $e\gamma$ 
collisions \cite{egamma}. In fact, the value of $\tanb$ can
be analytically determined by measuring parameters in the
gaugino sector of the MSSM, as outlined in \cite{lctanb}. 
Due to the clean experimental environment, 
stau pair production in $\epem$ collisions 
can be exploited to probe the SUSY
parameters, in particular $\tanb$, via left-right mixing \cite{Nojiri}.
Especially useful at high $\tanb$
will be measurements of the $\tau$ polarization
in $\widetilde\tau$ decays, which is directly sensitive to 
$m_\tau/(\cos\beta m_W)$~\cite{Boos:2002wu}.

Many SUSY particle production and decay processes
depend on $\sin\beta$ or $\cos\beta$ (and not their inverses). 
Once $\tanb\gsim 5$, $\sin\beta\sim 1$  and $\cos\beta$ is simply
small, and thus even large changes in $\tanb$ will have
little impact on the related experimental observables at large 
$\tanb$~\cite{egamma}. There are exceptions to
this statement, including the $\tau$ polarization measurement
referenced above. However, even this process
involves other SUSY parameters at tree-level simultaneously as well. 
Thus, while many of the SUSY parameters 
can be measured with high precision, the accuracy with
which $\tanb$ can be determined using measurements not involving
the Higgs bosons remains uncertain,
especially if $\tanb$ is large. Further, in non-SUSY extensions
of the SM, it may be that the only direct probe of $\tanb$ will
be via the Higgs sector. As seen from 
Eqs.~(\ref{equ:hff})--(\ref{equ:Htb}), the
heavy Higgs boson couplings to fermions are very sensitive to, 
and provide a direct probe of, $\tanb$ at tree-level. Pair production of
heavy Higgs bosons $AH$ and $H^+H^-$ was studied in $e^+e^-$
collisions \cite{hh,Gunion:1996cc,Gunion:1996qd},
and improved sensitivity to $\tanb$ was obtained
by considering  $H^\pm t\bar b$ \cite{feng} and
$A/Hb\bar b, A/Ht\bar t$ \cite{tao}. If a muon collider 
becomes available to produce a heavy Higgs boson in the
$s$-channel, its coupling 
could be measured with very excellent precision \cite{muon}.

In this paper, we perform a comprehensive analysis of $\tanb$
determination via heavy Higgs boson production and decay
 at an $e^+e^-$ linear collider with $\sqrt s=500$
GeV and an integrated luminosity of 2000 $\fbi$. 
We amplify upon the results for the heavy neutral Higgs bosons obtained during
the last Snowmass workshop (as summarized in \cite{earlier}) and
extend our study to include the charged Higgs boson.
We show how various 
Higgs boson measurements can be used to determine $\tanb$. The different
types of measurements we consider are complementary in that some provide good
precision at low $\tanb$ and others at high $\tanb$; combined, a good
determination of $\tanb$ is possible throughout
its whole range. We include background simulations
and realistic $b$-tagging efficiencies.
In Sec.~II, we focus on the heavy Higgs bremsstrahlung process
$\bb\ha\to \bb\bb$, the production rate for which is directly proportional
to $\tan^2\beta$. We then include the $\bb\hh\to\bb\bb$ process
in the context of the MSSM and estimate the accuracy with
which $\tanb$ can be determined by combining experimental
results for both processes. In Sec.~III, 
we examine the pair production of a CP-even
Higgs boson ($\hl$ or $\hh$) and the CP-odd $\ha$, followed by decay
of the Higgs bosons to the $\bb\bb$ final state.
In particular, at large $\tanb$,
the total decay widths of the heavy Higgs bosons can be broad since
these widths are proportional to $\tan^2\beta$. 
The resulting accuracy for the $\tanb$ determination is obtained.
We extend these studies to charged
Higgs in Sec.~IV. In Sec.~V, we briefly summarize the
LHC sensitivity to $\tanb$ deriving from heavy Higgs production.
Finally, we summarize our results in Sec.~VI.

Before proceeding with our analysis, we would like to 
point out that we are taking a phenomenological approach to the
$\tanb$ determination. Namely, we only consider $\tanb$ as an
effective way of specifying the coupling for the Higgs bosons and fermions 
through the usual tree-level relations
and explore the extent to which this coupling can be experimentally determined
at the linear collider experiments. After including radiative
corrections, the relation of $\tanb$ to the various Yukawa couplings becomes 
process-dependent. For a recent theoretical discussion on
the issue of the gauge dependence of these relations,
see Ref.~\cite{yamada-tanb}. In this context, our results
should be viewed as giving the accuracy with which the actual Yukawa couplings
can be measured.

\vspace*{-.6cm}
\section{The \boldmath $\bbA\to \bb\bb$ bremsstrahlung process}
\vspace*{-4mm}

Searches for $b\bar b\ha$ and $b\bar b \hl$
were performed in the four-jet channel using 
LEP data taken at the $Z$ resonance~\cite{l3,aleph,delphi,opal}.
In this section, we consider a linear collider with a center-of-mass energy
of 500 GeV or higher, and begin by focusing just 
on the $\bb\ha$ production process that probes
the direct coupling of the 
CP-odd Higgs $\ha$ to $b\bar b$. Our analysis will employ
cuts designed to eliminate Higgs pair resonant production, which, when
kinematically accessible, dominates
$\ha$ production before cuts but is less sensitive to $\tanb$.
The challenge of this study is the low expected production rate and the 
large irreducible background for a four-jet final state, as discussed 
in a previous study~\cite{epj}.
A LC analysis has been performed using event generators for 
the signal process $\epem\to\bb\ha\to\bb\bb$~\cite{generator} and the
$\epem \to eW\nu,~\epem Z,~WW,~ZZ,~q\anti q~(q=u,d,s,c,b),~t\anti t,~\hl\ha$ 
background processes~\cite{background} that
include initial-state radiation and beamstrahlung.

\begin{figure}[tb]
\begin{center}
\vspace*{-1.5cm}
%\mbox{\epsfig{file=purity.eps,width=0.6\textwidth}}
\includegraphics[width=0.5\textwidth]{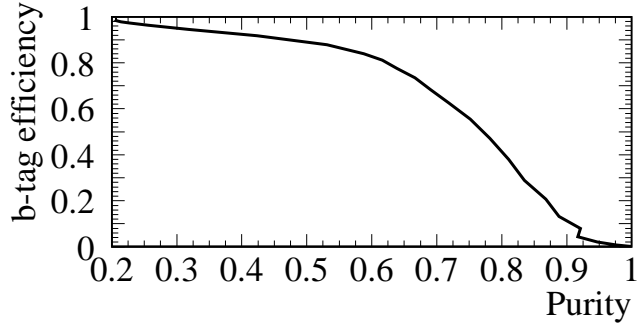}
\vspace*{-0.9cm}
\end{center}
\caption{\label{fig:effpurity} Simulated $b$-tagging performance at the TESLA.}
\end{figure}

For a 100 GeV CP-odd Higgs boson and $\tanb=50$,
the signal cross section is about 
2~fb \cite{maria,Djouadi:1991tk,Grzadkowski:1999ye}.
The generated events were passed through the fast detector
simulation SGV~\cite{sgv}.
The detector properties closely follow the TESLA detector 
Conceptual Design Report~\cite{tesla}.
%060700
The simulation of the $b$-tagging performance is very important
for this analysis.  
The efficiency versus purity distribution for the simulated
b-tagging performance is shown in Fig.~\ref{fig:effpurity} 
for the hadronic event sample $\epem\to q\bar{q}$ for
5 flavors, where efficiency is the ratio of 
simulated $\bb$ events with the selection cuts
to all simulated $\bb$ events,
and purity is the ratio
of simulated $\bb$ events with the selection cuts
to all selected $q\anti{q}$ events.
Details of the event selection and background reduction are 
described elsewhere~\cite{epj}.

%\newpage

For $\mha=100\gev$ in the 
context of the MSSM, the 
SM-like Higgs boson is the $\hh$ while the light $\hl$
is decoupled from $WW,ZZ$ [$\cos(\beta-\alpha)\sim 1$
and $\sin(\beta-\alpha)\sim 0$]. The $\bb\hl$ coupling
is essentially equal (in magnitude) to the $\bb\ha$ coupling ($\propto
\tanb$ at the tree level) and $\mhl\sim\mha$, implying that
the signal would be doubled from $\bb \ha$ and $\bb \hl$. 
 Also important will be $\hl \ha$ pair production, which is 
proportional to $\cos(\beta-\alpha)$ and will have full
strength in this particular situation; $\hh\ha$ production will be
strongly suppressed. We focus first on $b\anti b\ha\to\bb\bb$.

\begin{figure}[tb]
\vspace*{-.5cm}
\begin{minipage}{0.6\textwidth}
\begin{center}
\vspace*{-1.7cm}
\includegraphics[width=\textwidth]{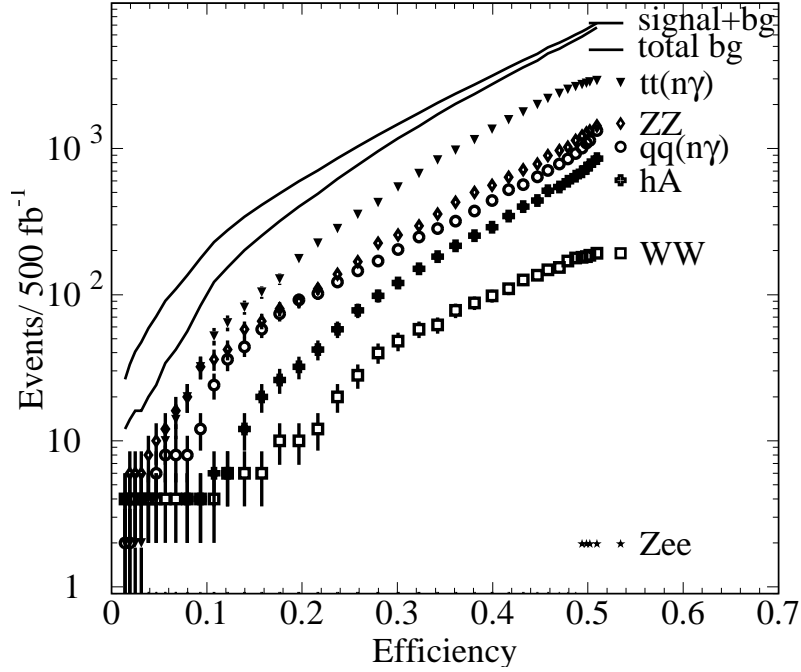}
\end{center}
\end{minipage}
\caption{\label{fig:ida2a}
Final background rate versus $\bb\ha$ signal efficiency for 
$\mha=100\gev$, $\rts=500\gev$ and $\call=500\fbi$. 
We take a fixed value of $m_b=4.62\gev$. 
}
\end{figure}

The expected background rate for a given $b\anti b\ha\to\bb\bb$ signal
efficiency is shown in Fig.~\ref{fig:ida2a}. 
One component of the background is $\hl\ha\to \bb\bb$ since it has
rather weak dependence on $\tanb$.
Our selection procedures are, in part, designed to reduce this piece of the
background as much as possible. Nonetheless, it may lead to 
significant systematic error in the determination of $\tanb$ 
due to interference with the signal, as discussed below. 
For the $\bb \ha\to\bb\bb$ signal, 
the sensitivity $S / \sqrt{B}$
for $\mha=100\gev$ is almost independent 
of the working point choice of signal efficiency in the range 
$\eps_{\rm sel}=5$\% to 50\%.
For a working point choice of 10\% efficiency, 
the total simulated background of about 
16 million events is reduced to 100 background events
with an equal number of signal events at $\tanb=50$.
We estimate the error on determining $\tanb$ by
\vspace*{-1mm}
\begin{equation}
\Delta\tan^2\beta / \tan^2\beta = \Delta {  S} / {  S}
=\sqrt{{  S} + {  B}} / {  S} .
\vspace*{-1mm}%
\end{equation}
If this were the only contributing process, then we would have
$\sqrt{{  S} + {  B}} / {  S} \approx 0.14$, 
resulting in an error on $\tanb= 50$ of $7\%$.
For smaller values of $\tanb$, the sensitivity decreases rapidly.
A $5\sigma$ signal detection is still possible for $\tanb= 35$.
In the MSSM context, the $\bb\hl$ signal would essentially double the
number of signal events and have exactly the same $\tanb$ dependence,
yielding $\Delta\tan^2\beta/\tan^2\beta\sim \sqrt{300}/200\sim 0.085$
for $\tanb=50$.

\begin{figure}[tb]
\begin{minipage}{0.6\textwidth}
\begin{center}
\vspace*{-1.7cm}
\includegraphics[width=\textwidth]{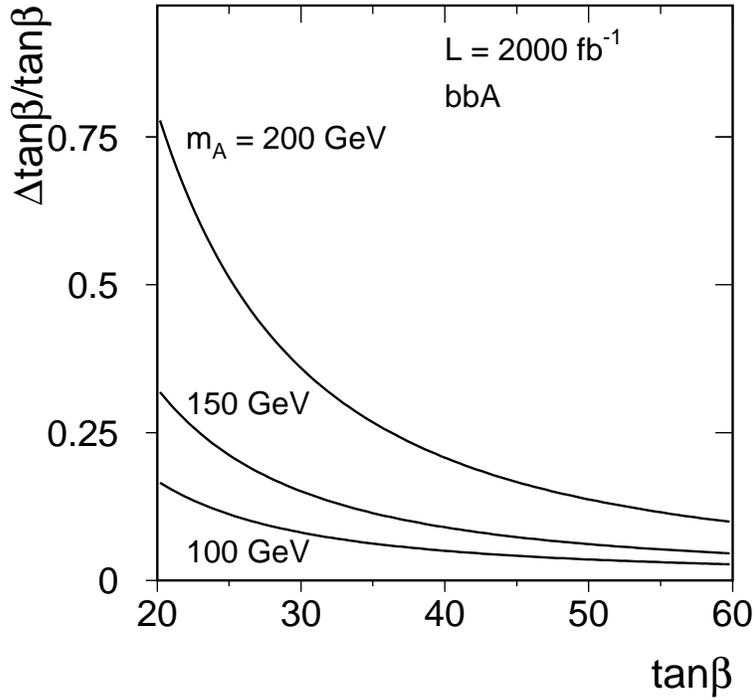}
\end{center}
\end{minipage}
\caption{\label{fig:ida2b}
The $\tanb$ statistical error for $\call=2000\fbi$
and $\mha=100,150,200\gev$ for 10\% selection efficiency.
For $\mha=100\gev$, the signal and 
background rates are four times those
given in Fig.~\ref{fig:ida2a} at the 10\% efficiency
point. Similar results are employed at $\mha=150$ and $200\gev$.}
\end{figure}

Although 
the number of $\hl\ha$ background events is very small compared to the
other background reactions 
after the event selection,
interference between the signal 
($ \bb\ha\to \bb\bb$ plus $\bb\hl\to\bb\bb$) and the background 
($ \hl \ha\to \bb\bb$) could be important. 
At the working point of $10\%$ signal efficiency, and
after applying the selection procedures, the expected rate for the latter
is $2\pm1$ events for $\call=500\fbi$.
To assess the effect of the interference, let us momentarily retain only 
the $\bb\ha$ signal and the $\hl\ha$ 
background. We first calculate the cross sections 
$ \sigma(\epem\to \bb\ha\to \bb\bb)$,
$ \sigma(\epem\to \hl\ha\to \bb\bb)$, and
$ \sigma(\epem\to \bb\ha+\hl\ha\to \bb\bb)$
with CompHEP~\cite{comphep} before selections.
We define the interference as
\begin{equation}
\sigma_{\rm interf}=\sigma_{\bb\ha+\hl\ha}-\sigma_{\bb\ha}-\sigma_{\hl\ha}.
\end{equation}
For the default value $m_{b}=4.62$~GeV, at $\tanb=50$ we obtain 
$\sigma_{\bb\ha}=1.83\pm0.01$~fb, 
$\sigma_{\hl\ha}=36.85\pm0.10$~fb, 
$\sigma_{\bb\ha+\hl\ha}=39.23\pm0.12$~fb, and thus
$\sigma_{\rm interf}=0.55\pm0.16$~fb.
We observe a constructive interference similar in size to the signal. Thus, 
more signal events are expected than simulated and 
the statistical error estimate is conservative.
After selection cuts, we have found 100 signal events versus two $\hl\ha$
background events.  The maximum interference magnitude arises if 
the interference events are signal-like, yielding an interference excess of
$(10+\sqrt 2)^2-100-2\sim 28$, a percentage ($\sim 30\%$) similar to
the ratio obtained before selection cuts. 
If the events from the interference are background-like,
the resulting systematic error will be small, since the $\hl\ha$ background 
is only a small part of the total background. 
In the MSSM context we have an exact prediction 
as a function of $\tanb$ for
the combined contribution of $\hl\ha\to \bb\bb$ and $\bb\ha\to\bb\bb$
(plus $\bb\hl\to\bb\bb$), including all interferences, 
and this exact prediction can be compared to the data.
In order to test this exact prediction, it may be helpful
to compare theory and experiment for several different event selection
procedures, including ones that give more emphasis to the $\hl\ha$ process.
Of course, this exact prediction depends somewhat on other
MSSM parameters, especially if decays of the $\hl$
or $\hh$ to pairs of supersymmetric particles are allowed or
ratios of certain MSSM parameters are relatively large~\cite{Carena:1998gk}. 
If this type of uncertainty exists,
the systematic error on $\tanb$ can still be controlled by 
simultaneously simulating all sources of $\bb\bb$ events for various 
$\tanb$ values and fitting the complete data set
(assuming that the other MSSM parameters are known sufficiently well). 
Another possible theoretical systematic uncertainty derives from 
higher-order corrections. The full next-to-leading order (NLO) QCD corrections 
are given in \cite{Dittmaier:mg,Reina:2001sf}. There it is
found that using the running $b$-quark mass incorporates
the bulk of the NLO corrections. For example,  for $\mha=100\gev$, employing
$m_b(100\gev)\sim 2.92\gev$ versus $m_b(m_b)\sim 4.62\gev$ yields (before cuts)
a cross section of $\sim 0.75\fb$ versus $\sim 2\fb$, 
respectively, at $\tanb=50$. 
The signal rates and resulting errors quoted in this section 
are those computed using $m_b=4.62\gev$. Use of the running mass would reduce
the event rates and increase our error estimates. In subsequent
sections and figures, all results and errors are computed in the MSSM
context using the running $b$-quark mass. 
Higher-order corrections of all kinds will be better known
by the time the Linear Collider (LC) 
is constructed and data is taken and thus should not be
a significant source of systematic uncertainty.
An experimental challenge is associated
with knowing the exact efficiency of the event selection procedure.
At the working point of an efficiency $\eps_{\rm sel}=10\%$, to achieve 
$\Delta\tanb/\tanb<0.05$ requires $\Delta\eps_{\rm sel}/\eps_{\rm sel}<0.1$,
equivalent to $\Delta\eps_{\rm sel}<1\%$.

In addition to the $\hl\ha$ Higgs boson background, two other Higgs boson
processes could lead to a $\bb\bb$ topology. 
First, the process $ \epem\to \hh Z$ can give a $\bb\bb$ final state.
In fact, for large $\tanb$ the $\hh Z$ cross section 
is maximal and similar in size to the $\hl\ha$ cross section. Nonetheless,
its contribution to the background is much smaller because
the $ \hh Z\to\bb\bb$ branching is below 10\% compared
to about 80\% for $\hl\ha\to\bb\bb$. 
Since the $\hl\ha$ process contributed only
2\% of the total background, the contribution to the background
from the $\hh Z$ process can be neglected.
The second Higgs boson process leading to a $\bb\bb$ topology is 
that already discussed, $ \epem\to\bb \hl$. The only distinction
between this and the $ \epem\to\bb \ha$ process
is a small difference in the angular distribution
due to the different production matrix elements. 
Thus, the selection efficiency is almost identical. The production rate
of the $ \bb \ha$ process is proportional to 
$\tan^2\beta$ while the $ \bb \hl$
production rate is proportional to $\sin^2\alpha / \cos^2\beta$.
In the MSSM context, this latter factor is $\sim\tan^2\beta$ for 
$\mha \ge 100\gev$ and large $\tanb$ (assuming $M_{\rm SUSY}\sim 1\tev$). 
In the general 2HDM, since $\tanb\approx 1/\cos\beta$ at large $\tanb$, 
the expected rate depends mostly on $\sin\alpha$ and the $\hl$ mass.
In this more general case where $\mhl \approx \mha$ but 
the MSSM expectation of $\alpha\sim -\beta\sim -\pi/2$ does not hold,
the enhancement of the $\bb\ha$ signal by
the $ \bb \hl$ addition would only allow a determination of
$|\sin\alpha|$ as a function of the presumed value of $\tanb$
(using the constraint that one must obtain the observed
number of $\bb\hl+\bb\ha$ events). Independent measurements
of the $\hh Z$ and $\hl\ha$
 production rates would then be needed to determine
the value of $\beta-\alpha$ and only then could
$\alpha$ and $\beta$ be measured separately.

It is essential for the $\tanb$ determination that 
a very high integrated luminosity can be accumulated
(we assume $\call=2000\fbi$ after several years
of data-taking). 
Fig.~\ref{fig:ida2b} shows the expected statistical error on $\tanb$ for 
$\mha = 100, 150$ and 200~GeV, assuming that the only measured
process is $\bb\ha$ with the help of our selection cuts.  
At the two higher $\mha$ values, in the MSSM context
it is the $\hh$ that would be decoupled and have mass
$\mhh\sim\mha$ and the $\hl$
would be SM-like.  Consequently, the $\bb\hh$ rate would be 
essentially identical to the $\bb\ha$ rate
and, assuming that one could verify the MSSM Higgs context by independent
means, would lead to still smaller $\tanb$ statistical errors than plotted,
the exact decrease depending upon the signal-to-background ratio.
For $\mha=150$ and $200\gev$, the $\hh\ha$ process (like the $\hl\ha$
process at $\mha=100\gev$) would have to
be computed in a specific model context or its relative weight
fitted by studying $\bb\bb$ production in greater detail in order
to minimize any systematic error from this source. Results obtained
in the case of the MSSM will be given in the following section.

\vspace*{-.6cm}
\section{\boldmath $\hh\ha$ production:
decay branching ratios and total widths}
\vspace*{-4mm}

The branching ratios
for $\hh$, $\ha$ and $\hpm$ decay to various allowed modes
vary rapidly with $\tanb$ in the MSSM when $\tanb$ is in the low to moderate
range, roughly below 20. 
Consequently, if these branching ratios can be measured accurately,
$\tanb$ can be determined with good precision in this range.
Measurement of the branching fractions is most easily accomplished 
using $\hh\ha$ and $\hp\hm$ pair production.
In particular, the pair production processes 
are nearly independent of $\tanb$ so that the rate in a given
channel provides a fairly direct probe of the branching ratio
for that channel. That $\tanb$ could be accurately determined 
using Higgs branching ratios measured in pair production 
was first demonstrated in~\cite{Gunion:1996cc,Gunion:1996qd}.
Refs.~\cite{Gunion:1996cc,Gunion:1996qd} consider 
a number of models for which SUSY decays of the $\hh$,
$\ha$ and $\hpm$ are kinematically allowed.
It was found that by measuring all available ratios of branching
ratios it was possible to determine $\tanb$ 
to better (often much better) than 10\% for $\tanb$ values 
ranging from 2  up to as high as 25 to 30
for $\mha$ in the 200--$400\gev$ range, assuming $\rts=1\tev$
and an effective luminosity (defined as the total luminosity times
the selection efficiency of the cuts
required to isolate the pair production process) of 
$\call_{\rm eff}=80\fbi$ (equivalent, for example, to $\call=2000\fbi$ for a
selection efficiency of 4\%).
A more recent analysis using a few specific points
in MSSM parameter space, focusing on the $\bb\bb$ event rate
and including a study at $\rts=500\gev$, is given in \cite{tao}.
This latter study uses a selection efficiency of 13\% 
and negligible background for detection of
$\epem \ra \hl\ha\to \bb\bb$ (relevant for $\mha\leq 100\gev$) 
or $\epem\to \hh\ha\to \bb\bb$ (relevant for $\mha\geq 150\gev$)
and finds small errors for $\tanb$ at lower $\tanb$ values.
Both \cite{Gunion:1996cc,Gunion:1996qd} and
\cite{tao} assume MSSM scenarios in which there are significant
decays of the $\ha$ and $\hh$ to pairs of SUSY particles, in particular
neutralinos and charginos.  These decays remain non-negligible
up to fairly high $\tanb$ values. As a result, the $\bb$ branching
fractions of the $\ha$ and $\hh$ increase more markedly as
$\tanb$ increases than if SUSY decays are absent.
Indeed, in the absence of SUSY decays, 
the $\bb\bb$ rate asymptotes quickly to a fixed value
as $\tanb$ increases. As we shall see, this means that much smaller errors
for the $\tanb$ determination using the $\hh\ha\to\bb\bb$ rate are achieved
if SUSY decays are present.

We now examine the errors on $\tanb$ that could be
achieved using Higgs pair production, following procedures related to those of 
\cite{Gunion:1996cc,Gunion:1996qd,tao}, but 
using updated luminosity expectations and somewhat more realistic
experimental assumptions and analysis techniques. 
We restrict the analysis to the process $\epem\to\hh\ha\to \bb\bb$,
ignoring possible additional sensitivity through ratios relative
to other final states. With both Higgs bosons reconstructed in their $\bb$
final state as two back-to-back clusters of similar mass,
backgrounds are expected to be negligible. 
All the results of this section are obtained using
version 2.0 of HDECAY \cite{hdecayref} for computing the branching
ratios and total widths of the Higgs bosons.

\begin{figure}[p]
\begin{minipage}{0.7\textwidth}
\begin{center}
\vspace*{-.7cm}
\hspace*{-.5in}
\includegraphics[width=.55\textwidth,angle=90]{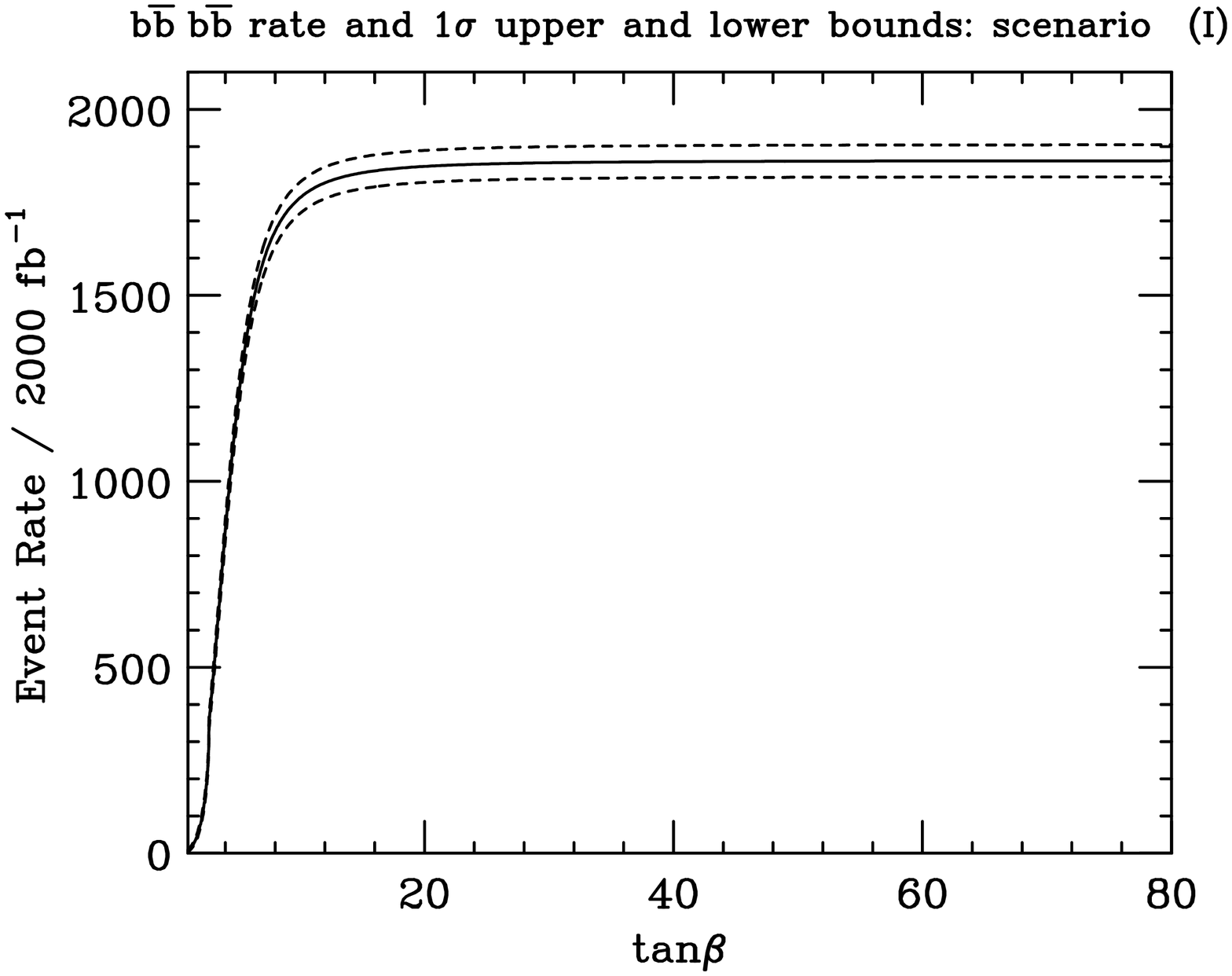}
\medskip
\hspace*{-.47in}
\includegraphics[width=.55\textwidth,angle=90]{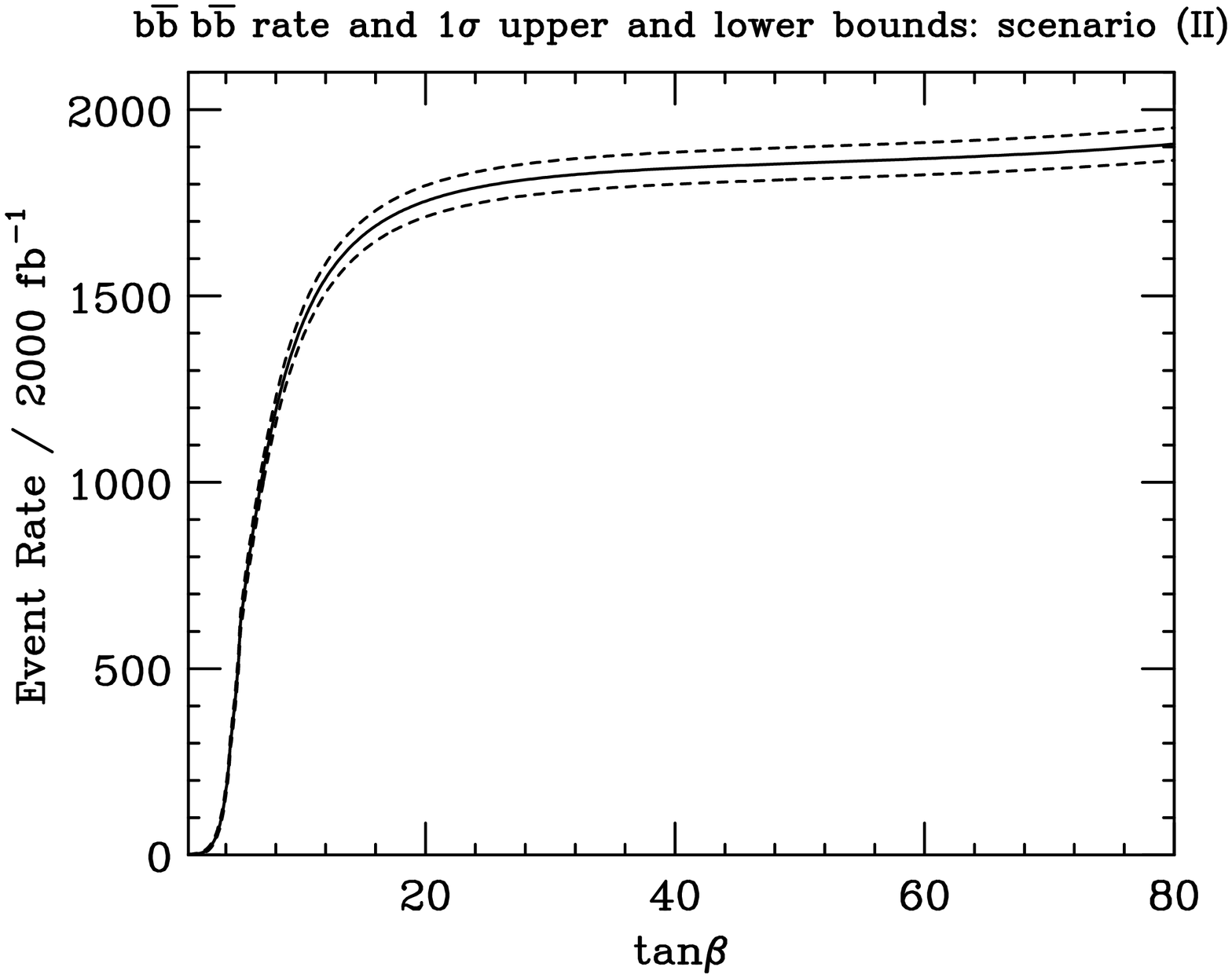}
\medskip
\hspace*{-.47in}
\includegraphics[width=.55\textwidth,angle=90]{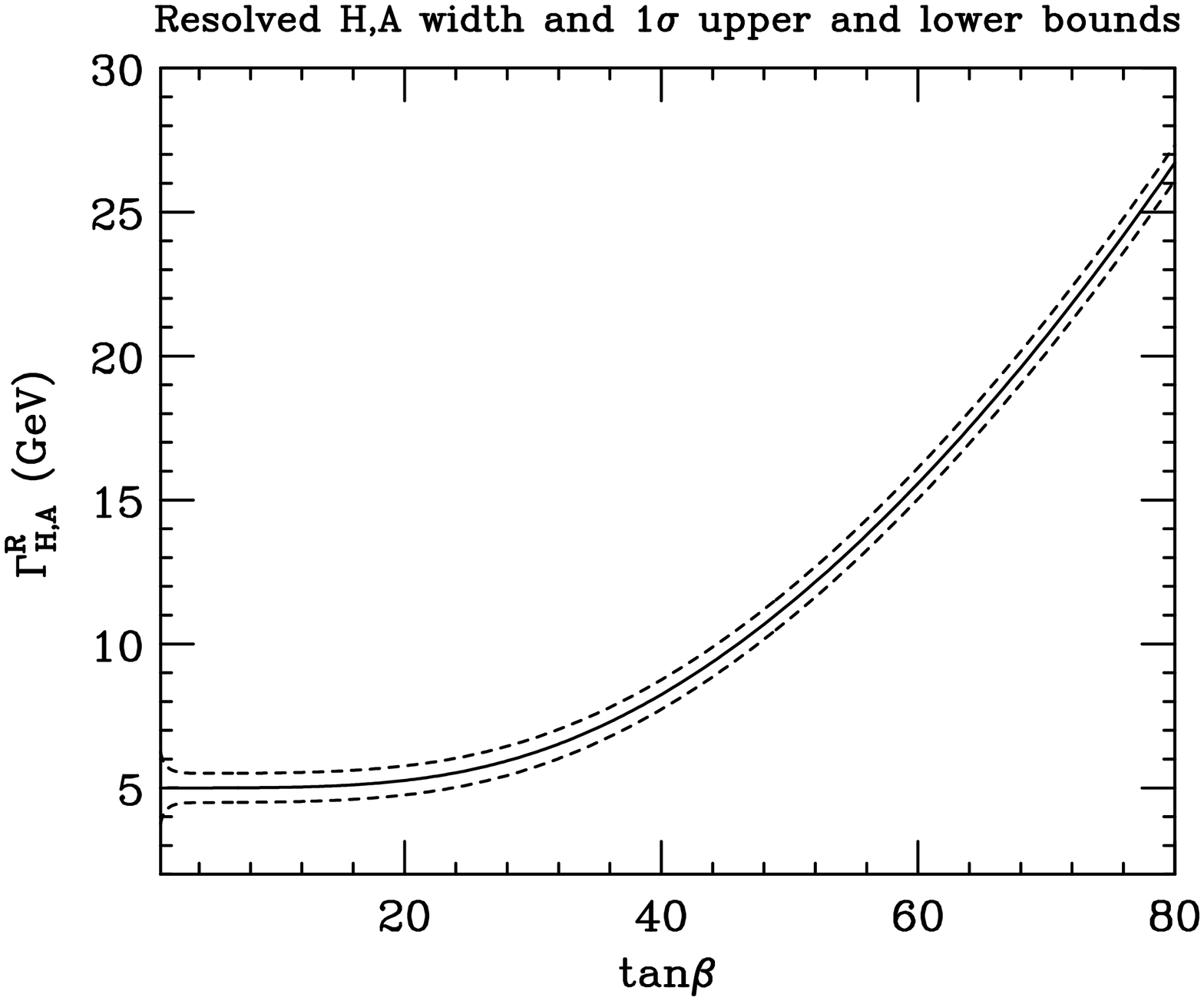}
\end{center}
\end{minipage}
\vspace*{-.2in}
\caption{\label{hhhawidrate}
Solid curves in the upper two figures give the rates  for
$\epem\to \hh\ha\to \bbbb$ in scenarios (I) and (II). The solid
curve in the lower figure is the resolved width
$\Gamma_{\hh,\ha}^{\rm R}$, Eq.~(\ref{gamr}), 
for scenario (I). Dashed curves in all three figures correspond
to the $1\sigma$ upper and lower bounds on these quantities.
We take $\mha=200\gev$, $\rts=500\gev$ and $\call=2000\fbi$.
An efficiency of 10\% is assumed
for cuts, acceptance and tagging.
The upper and lower $1\sigma$
bounds for $\Gamma_{\hpm}^{\rm R}$ include an additional
efficiency factor of 0.75 for keeping only events
in the central mass peak and assume the estimated mass resolution
of $\gamres=5\gev$, including 10\% systematic uncertainty.
Results for $\Gamma_{\hh,\ha}^{\rm R}$ in SUSY scenario case (II)
are very similar to those plotted for case (I). HDECAY \cite{hdecayref}
is used to compute the $\hh$ and $\ha$ widths and branching ratios.
}
\end{figure}

To understand the sensitivity to the presence of SUSY decays of the heavy
Higgs bosons, two different MSSM scenarios are considered:
\begin{description}
\itemsep=0.0in 
\item{(I)} $\mha=200\gev$, $m_{\wt g}=1\tev$, $\mu=M_2=250\gev$, \\ 
$m_{\wt t_L}=m_{\wt b_L}=m_{\wt t_R}=m_{\wt b_R}\equiv m_{\wt t}=1\tev$, \\ 
$A_\tau=A_b=0$, $A_t=\mu/\tanb+\sqrt6 m_{\wt t}$ (maximal mixing);
\item{(II)}  $\mha=200\gev$, 
$m_{\wt g}=350\gev$, $\mu=272\gev$, $M_2=120\gev$, \\ 
$m_{\wt t_L}=m_{\wt b_L}=356\gev$, $m_{\wt t_R}=273\gev$, 
$m_{\wt b_R}=400\gev$,\\
 $A_\tau=0$, $A_b=-672\gev$, $A_t=-369\gev$.
\end{description}
In scenario (I), SUSY decays of the $\hh$ and $\ha$ are kinematically
forbidden. Scenario (II) is taken from \cite{tao} in which
SUSY decays (mainly to $\wt \chi_1^0\wt\chi_1^0$) are allowed.
We will assume that appropriate event selection
criteria can be found such that for an event selection efficiency of 
$10\%$ there will be negligible background.  The resulting $\hh\ha\to\bb\bb$
event rates (per $2000\fbi$ of integrated luminosity) are plotted
for $\rts=500\gev$ 
in Fig.~\ref{hhhawidrate} as a function of $\tanb$.  The difference
in the dependence of the event rates on $\tanb$ is apparent.
In more detail: in scenario (I) 
the $\bb\bb$ event rates, after $10\%$ selection efficiency, 
are $8$, $77$, $464$, $1762$, and $1859$ 
at $\tanb=1$, $2$, $3$, $10$, and $40$, respectively. 
The corresponding event rates in scenario (II) are
$1$, $5$, $34$, $1415$ and $1842$.
These differing $\tanb$ dependencies imply significant 
sensitivity of the $\tanb$ errors to the scenario choice,
with worse errors for scenario (I). 
Finally, we note that for $\tanb>2$ the above event numbers
are such that backgrounds are indeed
negligible after 10\% efficient selection cuts; for $\tanb\sim 1$,
backgrounds might become an issue.

\begin{figure}[tb]
 \begin{minipage}{0.75\textwidth}
\begin{center}
\hspace*{-.8in}
%the old .ps name
%\includegraphics[width=1.0\textwidth,angle=90]{tanb_error_withsyserrorrunmb.ps}
\includegraphics[width=1.0\textwidth,angle=90]{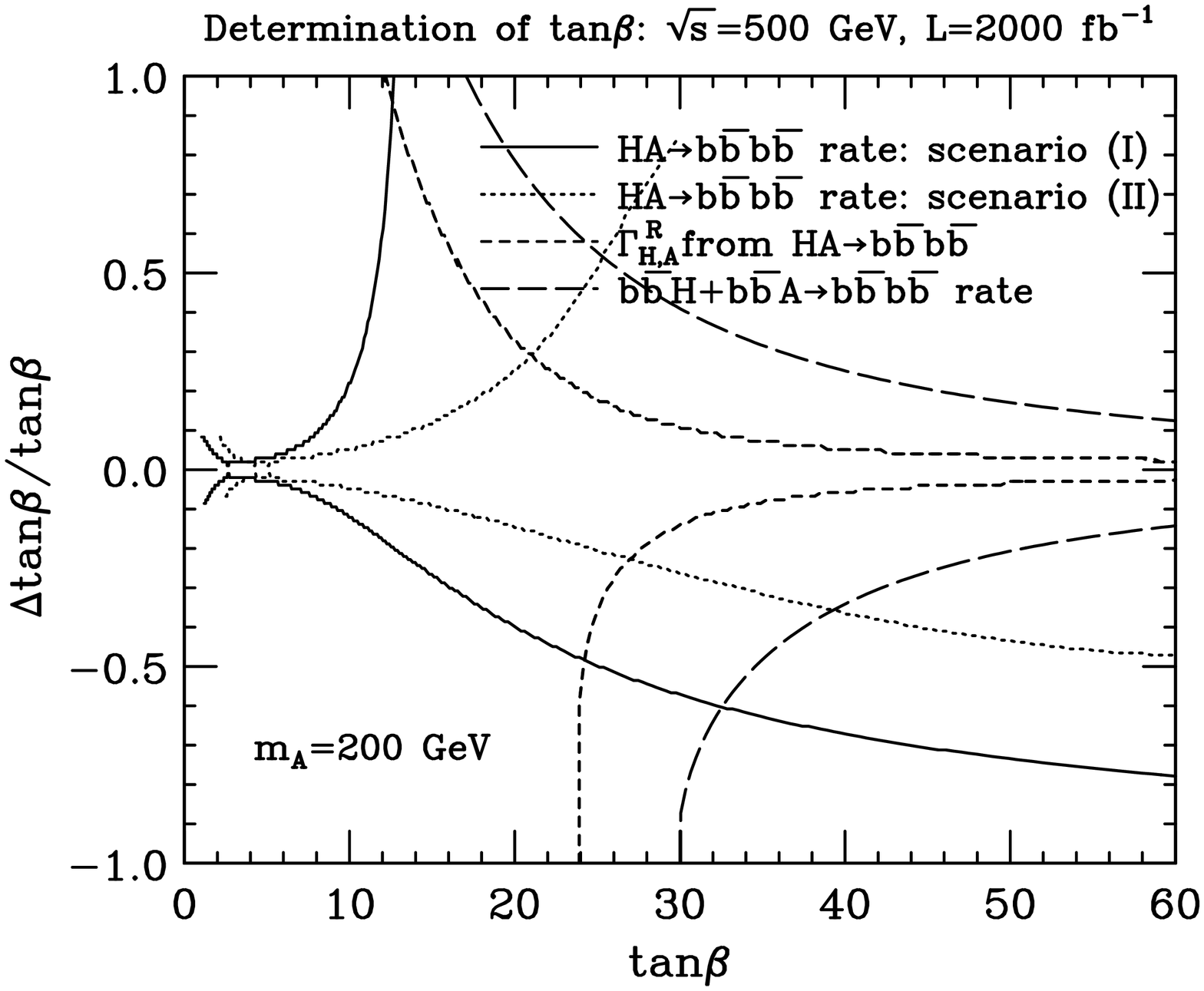}
\end{center}
\end{minipage}
\caption{\label{hhhaonly} 
For the MSSM with $\mha = 200$~GeV, and 
assuming $\call=2000\fbi$ at $\rts=500\gev$,
we plot the $1\sigma$ statistical upper and lower bounds, $\Delta\tanb/\tanb$,
as a function of $\tanb$ based on:
the rate for $\epem\to\bb\ha+\bb\hh\to\bbbb$;
the rate for $\epem\to\hh\ha\to\bbbb$; and
the average resolved width $\Gamma^{\rm R}_{\hh,\ha}$ 
defined in Eq.~(\ref{gamr}) for the $\hh$ and $\ha$ as determined
in $\epem\to\hh\ha\to\bb\bb$ events.
For the rates, results for SUSY scenarios (I) and (II) differ
significantly, as shown. For $\bb\ha+\bb\hh\to\bb\bb$
and $\Gamma^{\rm R}_{\hh,\ha}$ we show only the results
for MSSM scenario (I). Results for scenario (II) are essentially identical.
Upper and lower curves of a given type give the upper and lower $1\sigma$
bounds, respectively, obtained using a given process as shown in
the figure legend.
We include running $b$-quark mass effects and employ
HDECAY \protect\cite{hdecayref}.}
\end{figure}

To determine the $1\sigma$ 
statistical errors of the $\tanb$ determination, we compute,
for each choice of $\tanb$, the $1\sigma$ upper
and lower bounds on the expected event number as
$N(\bb\bb)\pm\sqrt{N(\bb\bb)}$. These upper and lower bounds
are also shown in Fig.~\ref{hhhawidrate}. The upper (lower) 
event rate numbers are required
to be $\geq 10$ to set an upper (lower) $\tanb$ bound, respectively.
Since the event number increases monotonically with $\tanb$
for both MSSM scenarios, we can
then use the given MSSM model scenario to determine the $\tanb$ value for which
the number of events is equal to the $1\sigma$ upper (lower) bound.
These values define the $1\sigma$ upper (lower) bound on $\tanb$, respectively.
The resulting fractional upper and lower limit errors $\Delta\tanb/\tanb$
are plotted for MSSM scenarios (I) and (II) in Fig.~\ref{hhhaonly}.
This procedure assumes that other measurements of SUSY
particle production at the LHC and the LC will have fixed the MSSM
scenario.

Let us discuss in more detail the $\tanb$ errors
from the $\hh\ha\to \bb\bb$ rate in scenario (II) as compared to 
scenario (I).  From Fig.~\ref{hhhawidrate} 
we see that in scenario (I) once $\tanb$ reaches 10 to 12
the $\bb\bb$ rate will not change much if
$\tanb$ is increased further since the $\hh\to \bb$ and $\ha\to\bb$
branching ratios approach constant values.
In contrast, if $\tanb$ is decreased the $\bb\bb$ rate declines significantly
as other decay channels come into play. Thus, meaningful lower bounds
on $\tanb$ are retained out to relatively substantial $\tanb$ values
whereas upper bounds on $\tanb$ disappear for $\tanb\gsim 10-12$.
In scenario (II), for reasons explained below, 
we have not plotted upper bounds  on $\tanb$
for $\tanb\gsim 30$. In fact, our numerical results indicate that
the upper bound on
$\Delta\tanb/\tanb$ decreases again as $\tanb$ increases
beyond 30.  We have traced this to the fact
that HDECAY predicts that $\mhh$ decreases (at fixed $\mha=200\gev$)
as $\tanb$ increases beyond 30.
This results in an increase of the $\hh\ha$ production cross section
with increasing $\tanb$. This, in turn, implies that 
the $\bb\bb$ rate increases (as shown in Fig.~\ref{hhhawidrate})
and that we can obtain an upper bound on
$\tanb$ despite the fact that the $\hh\ha\to\bb\bb$ final state branching
ratio approaches a constant value.
However, since this predicted decrease of
$\mhh$ at high $\tanb$ is somewhat peculiar to the precise
parameters chosen for scenario (II), we do not regard this result
as representative. For this reason, we have chosen not to 
show the scenario (II) upper limit curve beyond $\tan\beta=30$.
Had we plotted the region above $\tanb=30$, one would see a slowly
declining upper limit on $\tanb$.

The above results can be compared to the $\tanb$ determination based on
the $\bb\hh+\bb\ha\to\bb\bb$ rate using the procedures
of Sec.~II applied in the MSSM model context. 
For the computation of this rate, our calculation of
the $\bb\hh$ and $\bb\ha$ cross sections includes the dominant radiative
corrections as incorporated via $b$-quark mass
running starting with $m_b(m_b)=4.62\gev$. 
The $\hh$ and $\ha$ branching ratios and widths
are computed using HDECAY. Since there is
little sensitivity of this rate to the MSSM scenario (for
the high $\tanb$ values for which this means
of determining $\tanb$ is useful) we only present results
for scenario (I); where plotted,
errors for $\tanb$ from the $\bb\hh+\bb\ha\to\bb\bb$ rate are 
essentially independent of the MSSM scenario choice.
The errors on $\tanb$ resulting
from the rate for $\bb\hh+\bb\ha\to\bb\bb$ quickly become
far smaller than those based on $\hh\ha\to \bb\bb$ once $\tanb\gsim 30$.
This is illustrated in
Fig.~\ref{hhhaonly}, which compares the results for $\Delta\tanb/\tanb$
obtained using the $\epem\to\hh\ha\to \bb\bb$ rate to those based
on the $\bb\hh+\bb\ha\to \bb\bb$ rate. 
This comparison shows the
natural complementarity between these two techniques
for measuring $\tanb$. However, with these two techniques alone,
there is always a range of intermediate-size $\tanb$ values for which
a good determination of $\tanb$ is not possible.

This ``gap'' can be partly filled, and the error
on $\tanb$ at high $\tanb$ can be greatly reduced,
by using the intrinsic total
widths of the $\hh$ and $\ha$ to determine $\tanb$. 
However, it is only for $\tanb>10$ that the intrinsic widths 
can provide a $\tanb$ determination. This is because (a)
the widths are only $> 5\gev$
(the detector resolution discussed below) for $\tanb> 10$ and (b) 
the number of events in the $\bb\bb$ final state becomes maximal once
$\tanb>10$. 

We now discuss the experimental issues in determining the Higgs boson width. 
The expected precision of the SM Higgs boson width 
determination at the LHC and at a LC was studied in \cite{volker}. 
As described in~\cite{volker}, a simple estimate 
(based on a detector energy flow resolution of $\Delta E / E = 0.3 / \sqrt{E}$
for each of the two $b$-jets) yields an expected detector
resolution of $\gamres=5\gev$ for $\mha\sim 200\gev$.
However, an overall fit to the $b\anti b$ mass distribution
similar to the one in the study of \cite{volker} 
would give a Higgs boson resonance peak width 
%of $18.5\pm2.4$~GeV 
which is about $2\sigma$ larger than that expected from the convolution
of the 5 GeV resolution with the intrinsic Higgs width.
This can be traced to the fact that the overall fit includes
wings of the mass distribution that are present due to 
wrong pairings of the $b$-jets. The mass distribution contains
about 400 di-jet masses (2 entries per $\hh\ha$ event), of which about 
300 are in a central peak.  If one fits only the central peak,
the width is close to that expected based on
simply convoluting the 5 GeV resolution with the intrinsic Higgs width.
This indicates that about 25\% of the time wrong jet-pairings 
are made and contribute to the wings of the mass distribution.
Therefore, our estimates of the error on the
determination of the Higgs width will be based on the assumption that
only 3/4 of the events ({\it i.e.} those in the central peak)
retained after our basic event selection cuts (with assumed
selection efficiency of $10\%$)
can be used in the statistics computation.
The $m_{\bb}$ for each of the $\bb$
pairs identified with the $\hh$ or $\ha$ is binned in a single 
mass distribution. This is appropriate
since the $\hh$ and $\ha$ are highly degenerate for
the large $\tanb$ values being considered so that
the resolution of $5\gev$ is typically substantially
larger than the mass splitting. 
Our effective observable is then the resolved average width defined by:
\beq
\Gamma^{\rm R}_{\hh,\ha}=\half\left[\sqrt{[\gamhhtot]^2+[\gamres]^2}+
\sqrt{[\gamhatot]^2+[\gamres]^2}\right]\,.
\label{gamr}
\eeq
The resolved average width, $\Gamma^{\rm R}_{\hh,\ha}$,
for SUSY scenario (I) (including $\mha=200\gev$)
is plotted in Fig.~\ref{hhhawidrate}
as a function of $\tanb$. The results for scenario (II) are 
indistinguishable.

In order to extract the implied $\tanb$ bounds, 
we must account for the fact that
the detector resolution will not be precisely determined.
There will be a certain level of systematic uncertainty
which we have estimated at 10\% of $\Gamma_{\rm res}$, {\it i.e.} 
$\Delta\Gamma_{\rm res}^{\rm sys}=0.5 \gev$.
This systematic uncertainty considerably weakens our ability to
determine $\tanb$ at the lower values of $\tanb$ for which
$\gamhhtot$ and $\gamhatot$ are smaller than $\Gamma_{\rm res}$. This
systematic uncertainty
should be carefully studied as part of any eventual experimental analysis.
Given $\Gamma_{\rm res}$, $\Delta\Gamma_{\rm res}^{\rm sys}$ and
the number of selected $\bb\bb$ events, $N(\bbbb)$, we compute the useful
number of entries in the $\bbbb$ mass distribution 
for determining $\Gamma^{\rm R}_{\hh,\ha}$ as $N_{\rm entries}=
2\times 0.75\times N(\bbbb)$. The factor of 2 is 
because each $\bbbb$ event results in two entries, one for the $\hh$
and one for the $\ha$, and the factor of $0.75$ is that for retaining
only the central peak of the distribution.
The error for $\Gamma^{\rm R}_{\hh,\ha}$ is then computed
(following the procedure of \cite{Gunion:1996cn}) as
\beq\label{dgamr}
      \Delta\Gamma^{\rm R}_{\hh,\ha}=\left[
\left({\Gamma^{\rm R}_{\hh,\ha}\over \sqrt{2N_{\rm entries}}}\right)^2+
\left(\Delta\Gamma_{\rm res}^{\rm sys}\right)^2\right]^{1/2}\,.
\eeq
The resulting upper and lower $1\sigma$ bounds on $\Gamma^{\rm R}_{\hh,\ha}$
are plotted in Fig.~\ref{hhhawidrate}.  The upper and lower limits
on $\tanb$ are then obtained as the values $\tanb\pm\Delta\tanb$ for which
the central prediction (the solid curve of Fig.~\ref{hhhawidrate})
agrees with the values 
$\Gamma^{\rm R}_{\hh,\ha}\pm \Delta\Gamma^{\rm R}_{\hh,\ha}$. 
In computing $\Delta\Gamma^{\rm R}_{\hh,\ha}$ we have assumed a selection
efficiency of 10\% for computing $N(\bbbb)$. These errors are for
$\call=2000\fbi$ and $\rts=500\gev$. That an excellent
determination of $\tanb$ will be possible at high $\tanb$ is apparent.
The resulting accuracy for $\tanb$ obtained from measuring the average
(resolved) $\hh/\ha$ width is shown in Fig.~\ref{hhhaonly}. We see that good
accuracy is already achieved for $\tanb$ as low as 25 with
extraordinary accuracy predicted for very large $\tanb$.  The sharp
deterioration in the lower bound on $\tanb$ for $\tanb\lsim 24$ occurs
because the width falls below $\Gamma_{\rm res}$ as $\tanb$ is taken
below the input value and sensitivity to $\tanb$ is lost.  If there
were no systematic error in $\Gamma_{\rm res}$, this sharp fall off
would occur instead at $\tanb\lsim 14$.  To understand these effects
in a bit more detail, we again give some numbers for scenario (II). At
$\tanb=50$, $55$ and $60$, 
$\vev{\gamhhtot,\gamhatot}\sim 10.4$, $12.5$ and $14.9\gev$, 
respectively. After including the detector resolution, 
the effective average widths
become 11.5, 13.4 and 15.7 GeV, respectively, whereas the total
error in the measurement of the average width, including
systematic error, is $\sim 0.54 \gev$.  Therefore,
$\tanb$ can be determined to about $\pm 1$, or
to better than $\pm 2\%$. This high-$\tanb$ situation can
be contrasted with $\tanb=15$ and 20,
for which $\vev{\gamhhtot,\gamhatot}=0.935$ and 1.64 GeV, respectively,
which become 5.09 and 5.26 GeV after including detector resolution.
Meanwhile, the total error, 
including the statistical error and the 
systematic uncertainty for $\Gamma_{\rm res}$,
is about 0.57 GeV and no sensitivity to $\tanb$ is obtained.

The accuracies from the
width measurement are somewhat better than those
achieved using the $\bb\ha+\bb\hh\to\bb\bb$ rate measurement.
However, both of these high-$\tanb$ 
methods for determining $\tanb$ are important because they are
beautifully complementary in that they rely on very different
experimental observables. Further, both methods are nicely complementary in
their $\tanb$ coverage to the $\tanb$ determination based on the
$\hh\ha\to\bb\bb$ rate, which comes in at lower $\tanb$. 
In fact, the width measurement can provide a decent $\tanb$
determination even in the previously identified ``gap'' region where neither
the $\hh\ha\to\bb\bb$ nor the $\bb\ha+\bb\hh\to\bb\bb$
rates were able to provide such a determination. 
In particular, in the case of MSSM scenario (II), combining
the $\hh\ha\to\bb\bb$ rate and the width measurements implies that
the worst $\tanb$ lower bound is $\Delta\tanb/\tanb\sim -0.25$
at $\tanb\sim 28$ and the worst $\tanb$
upper bound is $\Delta\tanb/\tanb\sim +0.30$ at $\tanb\sim 20$.
However, in the case of MSSM scenario (I) a good upper bound  on $\tanb$ 
is not possible if $\tanb\sim 12-15$, even after including the width
measurement. Overall, there
is a window,  $10\lsim \tanb\lsim25$ in scenario (I) or
$20\lsim \tanb\lsim 25$ in scenario (II), for which an accurate 
determination of $\tanb$ ($\Delta\tanb/\tanb<0.2$) 
using just the $\bb\bb$ final state processes will not be possible.  
This window expands rapidly as $\mha$ increases (keeping $\rts$ fixed).
Indeed, as $\mha$ increases above $250\gev$, $\hh\ha$ pair production
becomes kinematically forbidden at $\rts=500\gev$ and, in addition,
 detection of the
$\bb\hh+\bb\ha$ processes at the LC (or the LHC) requires
\cite{Grzadkowski:1999wj} increasingly large values of $\tanb$.  This
difficulty persists even for $\rts\sim 1\tev$ and above; if
$\mha>\rts /2$, the $\hh$ and $\ha$ cannot be pair-produced and yet
the rate for $\bb\hh+\bb\ha$ production is undetectably small 
for moderate $\tanb$ values.

In the above study, we have not made use of other decay channels of
the $\hh$ and $\ha$, such as $\hh\to WW,ZZ$, $\hh\to \hl\hl$, $\ha\to
Z\hl$ and $\hh,\ha\to$~SUSY.  The theoretical studies of
\cite{Gunion:1996cc,Gunion:1996qd} indicate that their inclusion could
improve the precision with which $\tanb$ 
is measured at low to moderate $\tanb$
values.  A determination of $\Gamma^{\rm R}_{\hh,\ha}$ is also
possible using the $\bb\ha+\bb\hh\to\bb\bb$ events.  
To estimate how well $\tanb$ can be determined in this way,
let us assume that
$50\%$ of the events selected in the analysis of Sec.~II can be used
for a fit of the average width and that (as in the
$\hh\ha\to\bbbb$ study) a resolution of $5\gev$
can be achieved, based on a detector
energy flow resolution of $\Delta E / E = 0.3 / \sqrt{E}$. 
If we again assume that $10\%$
systematic error for the width measurement can be achieved, the
resulting $\tanb$ errors would be similar to
those obtained from the
$\bb\ha+\bb\hh\to\bb\bb$ event rate for $\tanb>30$ (see Fig.~\ref{hhhaonly}), 
{\it i.e.} not as small as the $\tanb$ errors obtained from the measurement of 
$\Gamma^{\rm R}_{\hh,\ha}$ in $\hh\ha\to \bbbb$ events.
A complete analysis that takes into account the significant background and the
broad energy spectrum of the radiated $\hh$ and $\ha$ is needed
to reliably assess the $\tanb$ errors that would be obtained
by measuring $\Gamma^{\rm R}_{\hh,\ha}$
in $\bb\ha+\bb\hh\to\bb\bb$ events. If $\tanb$ is large enough,
$\bb\ha+\bb\hh\to\bbbb$ would be observable in the MSSM
even when $\mha>\rts/2$, or, for a more general model, 
whenever $\hh\ha$ pair production is kinematically
forbidden but $\bb\ha\to\bbbb$ and/or $\bb\hh\to\bbbb$ production is allowed.
Then, the event rate for, and width measurements from $\bb\ha$ and/or $\bb\hh$
production would allow a determination of $\tanb$.  We have
not attempted a quantitative study of this situation.

Let us briefly return to the interpretation of these measurements
in terms of $\tanb$.  As stated in the introduction, we are
using $\tanb$ as a tree-level mnemonic to characterize
the $\bb$ Yukawa coupling of the Higgs bosons. For the 
soft-SUSY-breaking parameters for MSSM scenarios (I)
and (II), the one-loop corrections to 
the $b\anti b$ couplings of the $\hh$ and $\ha$ and the 
stop/sbottom mixing present in the one-loop corrections to the Higgs
mass matrix \cite{Carena:1998gk} are small.  More generally,
however, substantial ambiguity can arise, especially
if the sign and magnitude of $\mu$ is not fixed.
Assuming that these parameters are known,
the errors for the Yukawa coupling obtained from these 
measurements can be related to any given definition
of $\tanb$ and, except in very unusual cases, the resulting
error on $\tanb$ would be fairly insensitive to the precise scenario.
For example, one possible definition of $\tanb$ would be that
the $\mu^+\mu^-\ha$ coupling should be precisely given by $-(m_b/v)\tanb$,
see Eq~(\ref{equ:Htb}). This is a convenient definition
since the $\mu^+\mu^-\ha$ coupling will have very
modest higher-order corrections relative to the tree-level and
any such corrections can then be sensibly absorbed using the above
definition of $\tanb$. Given this definition of $\tanb$, 
the $\hh\bb$ and $\ha\bb$ couplings can be computed to any desired
order once the necessary MSSM parameters are known.  In this way,
all the probes of heavy Higgs Yukawa couplings discussed in this paper
can be related to this common definition of $\tanb$.

%*----------------------------------------
%{\bf NEW TEXT on H+H- to be placed near the end of the article.}

\section{ \boldmath$H^+H^-$ Production: decay branching ratios and total width}

In this section, we extend our study to include charged Higgs boson
production processes. Existing analyses
of $\epem\to\hp\hm$ production indicate that the absolute
event rates and ratios of branching ratios in various $\hp\hm$
final state channels will
allow a relatively accurate determination of $\tanb$ at low $\tanb$
\cite{Gunion:1996cc,Gunion:1996qd}. 
The process $\epem\to H^\pm tb$ can also be sensitive to 
$\tanb$ \cite{feng}. Here, we focus on
an experimentally based analysis of the
determination of $\tanb$ using the $\hp\hm\to\tbtb$ event rate.
As anticipated on the basis of the earlier work referenced above, we find that 
good accuracy can be achieved at low $\tanb$. We also demonstrate
that the total width of the $\hpm$ measured in the $tb$ decay channel
using $\hp\hm\to\tbtb$ production
will allow a fairly precise determination of $\tanb$ at high $\tanb$.
Since these two techniques for determining $\tanb$ are statistically 
independent of one another and
of the $\tanb$ measurements that employ neutral Higgs production,
they will increase the overall accuracy with which $\tanb$
can be measured at both low and high $\tanb$.

The reaction $\epem\to H^+H^-\to \tbtb$ can be 
observed at a LC~\cite{as:hphm500}, and recent high-luminosity 
simulations~\cite{lc800:kiiskinen} show that the cross section times
branching ratio can be measured precisely. As soon as the charged
Higgs boson decay into $tb$ is allowed this decay mode is dominant.
Nonetheless, $\br(\hpm\to tb)$ 
varies significantly with $\tan\beta$, especially for
small values of $\tanb$ where the $tb$ mode competes with
the $\tau\nu$ mode.
The $H^+\to t\anti b$ branching ratio and width are sensitive to $\tanb$
in the form
\be
\Gamma(\hpm\to tb) \propto \mt^2 \cot^2\beta + \mb^2\tan^2\beta\,.
\ee
As in Sec.~III, we will use HDECAY (which incorporates running
of the $b$-quark mass) to evaluate the charged Higgs boson
branching ratios and decay width.
It is useful to note that
the above form results in a minimum in the $tb$ partial width and branching
ratio in the vicinity of $\tanb\sim 6-8$. The depth of the minimum
in the branching ratio 
depends upon the extent to which the $tb$ channel is competing
against other modes.
In contrast, the cross section for $\epem\to\hp\hm$ production
is independent of $\tan\beta$ at tree-level. (The one-loop
corrections~\cite{guasch} result in a $10\%$ variation 
of the cross section with $\tanb$ 
which must be taken into account when the data are taken; we do not
include them in our study.)
The net result is that the rate for $\epem\to \hp\hm\to \tbtb$ 
has significant dependence
upon $\tanb$, coming mainly from the variation in the branching ratio.

\begin{figure}[p]
\vspace*{-.5cm}
\begin{minipage}{0.7\textwidth}
\begin{center}
\hspace*{-.5in}
\includegraphics[width=.55\textwidth,angle=90]{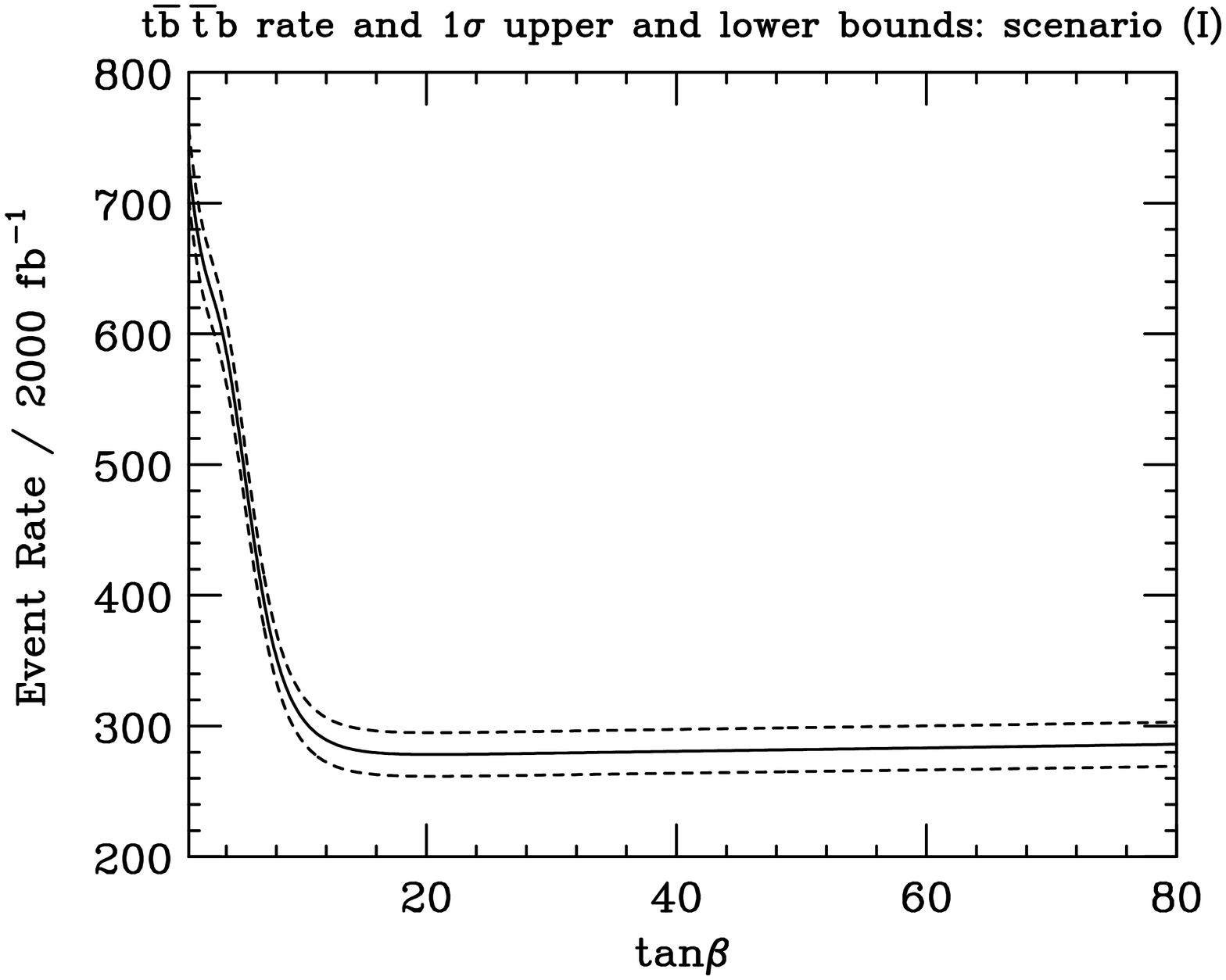}
\medskip
\hspace*{-.47in}
\includegraphics[width=.55\textwidth,angle=90]{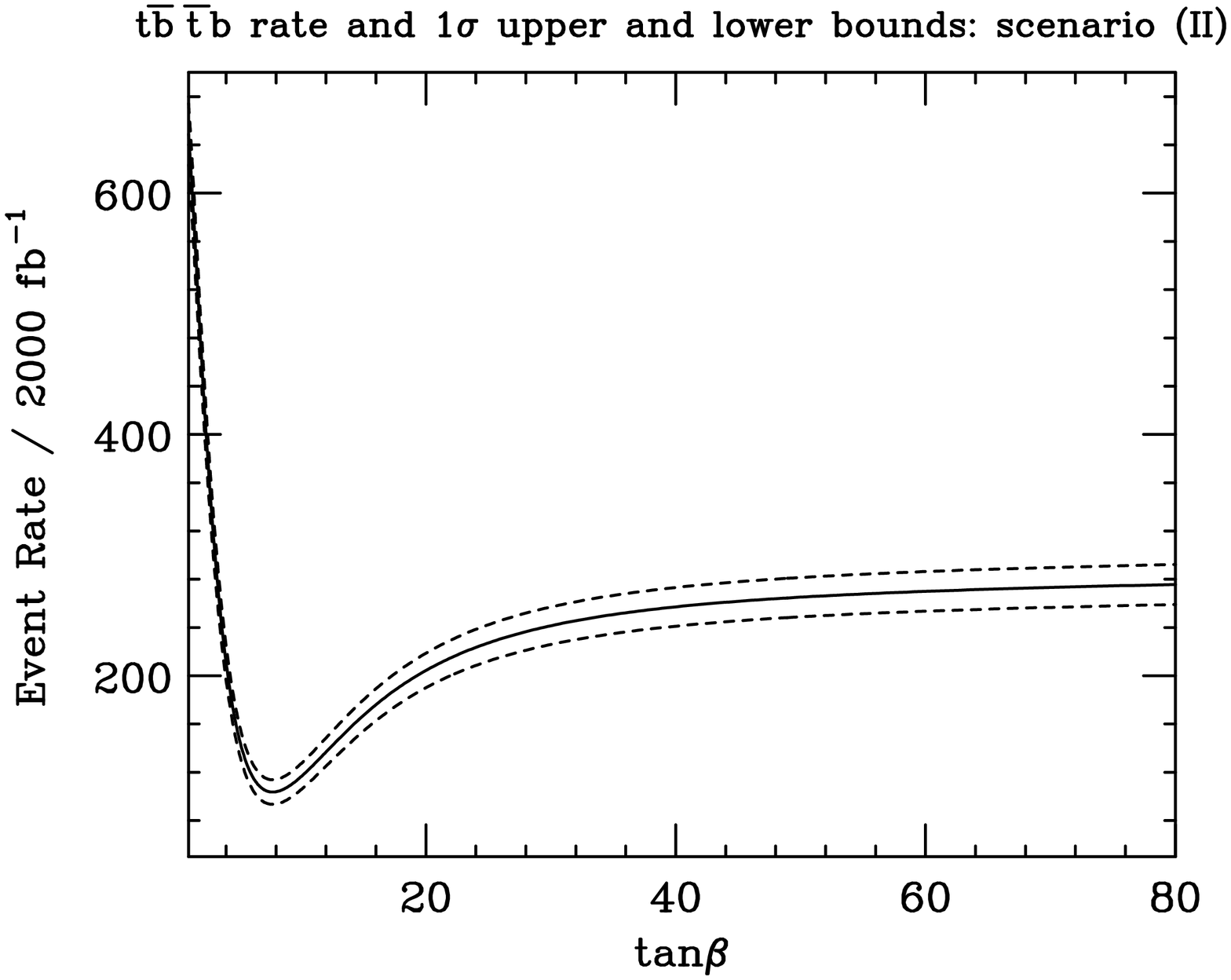}
\medskip
\hspace*{-.47in}
\includegraphics[width=.55\textwidth,angle=90]{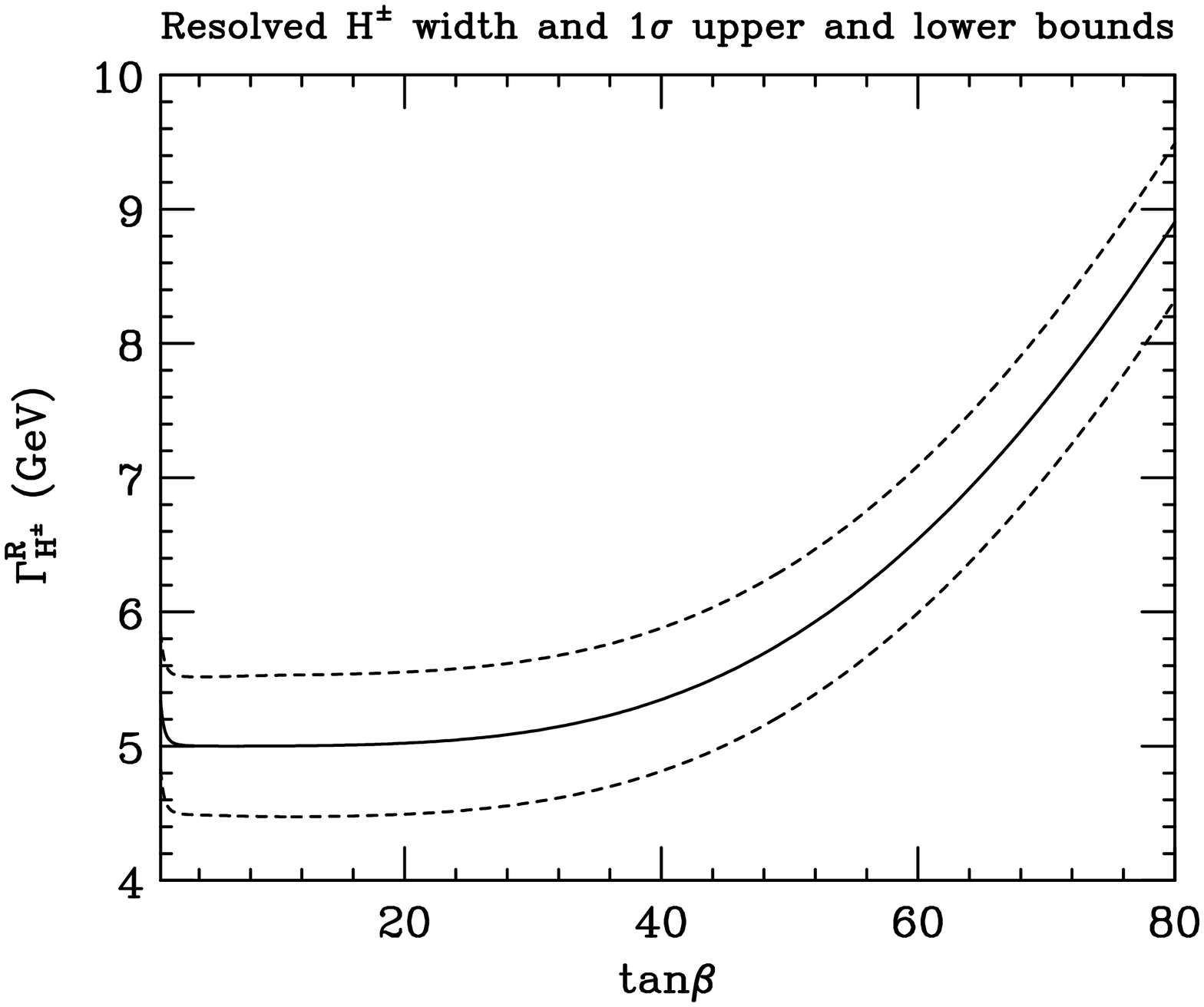}
\end{center}
\end{minipage}
\vspace*{-.2in}
\caption{\label{hpmwidrate}
The solid curves give the rates for
$\epem\to \hp\hm\to \tbtb$  for MSSM scenarios (I) and (II) in the upper two
figures and 
$\Gamma_{\hpm}^{\rm R}$, Eq.~(\ref{gamrhpm}), for scenario (I)
in the lower figure. The dashed curves are
the corresponding $1\sigma$ upper and lower bounds.
We take $\mhpm=200\gev$, $\rts=500\gev$ and $\call=2000\fbi$.
An efficiency of 2.2\% is assumed for cuts, acceptance and tagging. 
The upper and lower $1\sigma$
bounds  for $\Gamma_{\hpm}^{\rm R}$ include an additional
efficiency factor of 0.75 (which corresponds to keeping only events
in the central mass peak) and assume the estimated mass resolution
of $\gamres=5\gev$, including 10\% systematic uncertainty.
The $\Gamma_{\hpm}^{\rm R}$ results obtained in scenario (II)
are very similar to those plotted for scenario (I).
}
\end{figure}

Our procedures for estimating errors for the $\tbtb$
rate and for the total width 
are similar to those given earlier for $\hh\ha$ production
rates and width in the $\bbbb$ channel.
We base our efficiency for the $\tbtb$ final state on the study
of \cite{lc800:kiiskinen}.  For $\mhpm=300\gev$ at
$\rts=800\gev$, this study finds that the $\tbtb$ final state can
be isolated with an efficiency of $2.2\%$.
The reason for the much
lower efficiency as compared to 10\% efficiency for
the $\bbbb$ final state of $\hh\ha$
production is the difficulty of assigning the non-$b$ jets from $t$-decays 
to the correct top mass cluster.
For $\rts=500\gev$ and $\mhpm=200\gev$,  
we have adopted the same $2.2\%$ efficiency,
for which we assume little or no background
after cuts. For the total width determination, we assume that we
keep only 75\% of the events after cuts (\ie\
a fraction $0.75\times 0.022$ of the raw event number), corresponding
to throwing away wings to the mass peaks, and each $\tbtb$ event
is counted twice since we can look at both the $\hp$ and the $\hm$
decay.  We define a resolved width which
incorporates the intrinsic resolution for the width determination,
taken to be $\gamres=5\gev$:
\beq
\Gamma^{\rm R}_{\hpm}=\sqrt{[\gamhpmtot]^2+[\gamres]^2}\,.
\label{gamrhpm}
\eeq
Estimated errors based on the width measurement will assume a 10\% systematic
error in our knowledge of $\gamres$, {\it i.e.} 
$\Delta\gamres^{\rm sys}=0.5\gev$ as for the $\hh,\ha$ case.
We employ Eq.~(\ref{dgamr}), with the replacement
$\Gamma^{\rm R}_{\hh,\ha}\to\Gamma^{\rm R}_{\hpm}$,
to compute $\Delta\Gamma^{\rm R}_{\hpm}$.
In this case, $N_{\rm entries}=2\times 0.75\times N(\tbtb)$,
where $N(\tbtb)$ is computed using the above-noted selection
efficiency of $0.022$.
Figure~\ref{hpmwidrate} shows the resulting $\tbtb$ final state rate,
$N(\tbtb)$, for MSSM scenarios (I) and (II) and the resolved 
width $(\Gamma^{\rm R}_{\hpm})$ for scenario (I).
Also shown are the corresponding $1\sigma$ upper and lower 
bounds  on the rate and resolved width. 
These are then used in exactly the same manner as described
in the $\hh\ha$ case to determine the upper and lower bounds on $\tanb$.

For the rate, we observe from Fig.~\ref{hpmwidrate} that upper bounds
on $\tanb$ will be poor once $\tanb\gsim 10$ because of the
very slow variation of the $\tbtb$ final rate in this region.
At high $\tanb$, for SUSY scenario (I) 
lower bounds will be determined by the part
of the $\tbtb$ rate curve that rises rapidly when $\tanb$ falls below 10.
For SUSY scenario (II), the beginning of the dip will fix the
lower bounds on $\tanb$ when $\tanb$ is large, assuming that 
we know ahead of time from other experimental data that $\tanb$ is
larger than 15. As regards the width, 
the main point to note is that $\Gamma_{\hpm}^{\rm tot}$
rises only slowly with increasing $\tanb$. As a result,
the $5\gev$ resolution and 10\% systematic error for this resolution
are significant compared to the $< 10\gev$ $\hpm$ width that applies
throughout the $\tanb$ range studied. Note also, that for moderate $\tanb$
values, there will be no lower bound on $\tanb$ as a result
of the fact that $\Gamma^{\rm R}_{\hpm}$ never falls below $\gamres=5\gev$,
while the $1\sigma$ errors are substantially lower than this.
We will also assume that if $\tanb$ is large, then we will
know from other experimental information
(such as the $\hh\ha$ final state) that $\tanb$ is not small and that
the small rise in the width for $\tanb\sim 1$ is not relevant.

\begin{figure}[tb]
 \begin{minipage}{0.75\textwidth}
\begin{center}
\hspace*{-.8in}
\includegraphics[width=1.0\textwidth,angle=90]{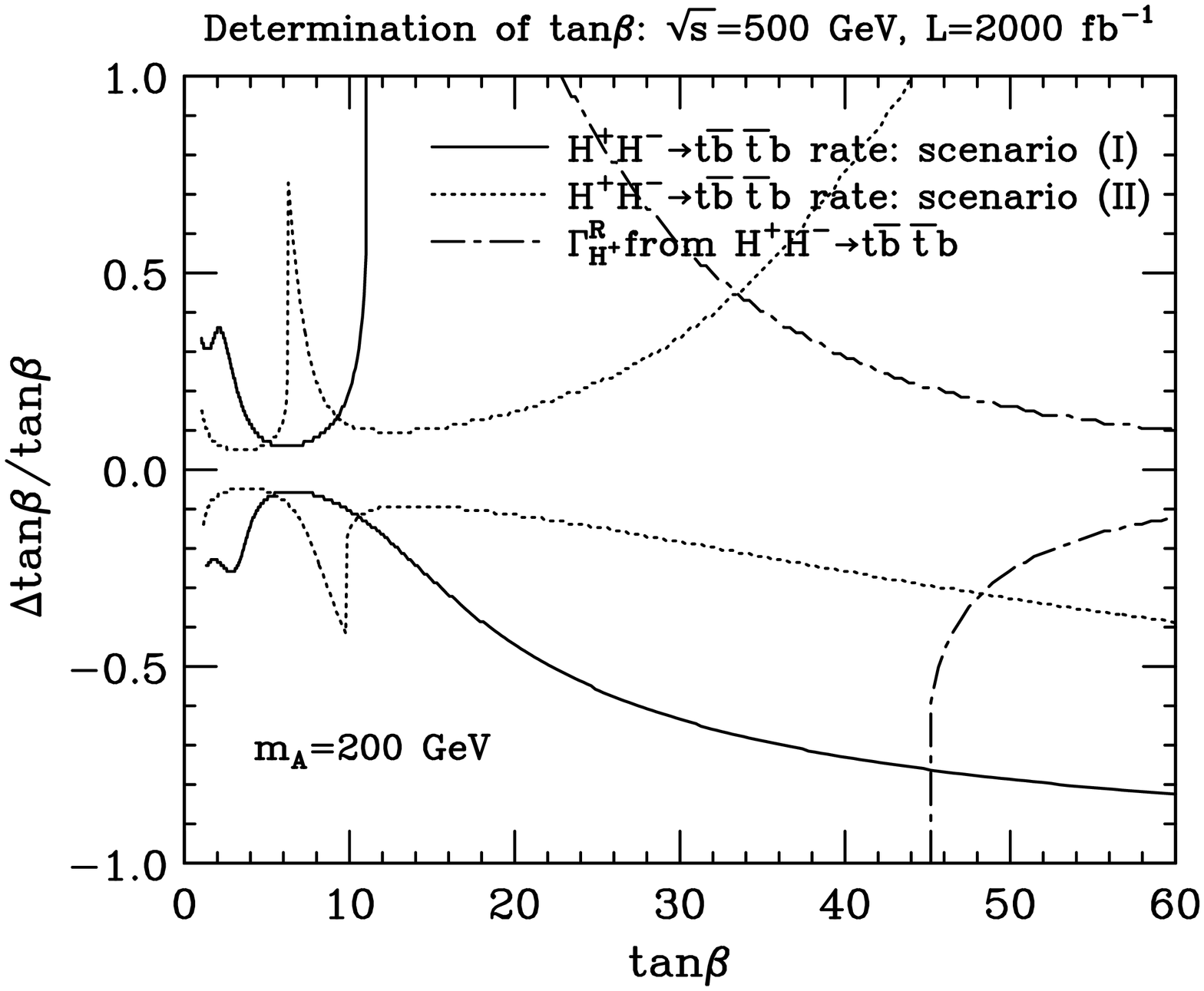}
\end{center}
\end{minipage}
\caption{\label{hpmonly} 
For the MSSM with $\mhpm \sim\mha= 200$~GeV, and 
assuming $\call=2000\fbi$ at $\rts=500\gev$,
we plot the $1\sigma$ statistical upper and lower bounds, $\Delta\tanb/\tanb$,
as a function of $\tanb$ based on:
the rate for $\epem\to\hp\hm\to\tbtb$; and
the resolved width $\Gamma^{\rm R}_{\hpm}$ 
defined in Eq.~(\ref{gamrhpm}) as determined
in $\epem\to\hp\hm\to\tbtb$ events.
For the rates, results for SUSY scenarios (I) and (II) differ
significantly, as shown. For 
$\Gamma^{\rm R}_{\hpm}$ we show only the results
for MSSM scenario (I). Results for scenario (II) are essentially identical.
Upper and lower curves of a given type give the upper and lower $1\sigma$
bounds,  respectively, obtained using a given process as shown in
the figure legend.
We include running $b$-quark mass effects and employ
HDECAY \protect\cite{hdecayref}.}
\end{figure}

The resulting $\tanb$ upper and lower bounds
appear in Fig.~\ref{hpmonly}. Comparing to Fig.~\ref{hhhaonly},
we observe that for SUSY scenario (I) the $\tbtb$ rate measurement gives a 
$\tanb$ determination that is quite competitive with that from $\hh\ha$
production in the $\bbbb$ final state.  For SUSY scenario (II),
the $\tbtb$ rate gives an even better $\tanb$ determination
than does the $\bbbb$ rate. On the other hand,
the width measurement from the $\tbtb$ final state of $\hp\hm$
production is much poorer
than that from the $\bbbb$ final state of $\hh\ha$ production,
as was to be expected from the discussion given earlier.

The rate for $\epem\to
t\anti b H^-+\anti t b H^+ \to t\anti t b\anti b$ is also very
sensitive to $\tanb$ and might be a valuable addition to the
$\epem\to\hp\hm\to \tbtb$ and 
$\epem\to\bb\ha+\bb\hh\to \bb\bb$ rate determinations of $\tanb$.  The 
theoretical study of
\cite{feng} finds, for example, that if $\mhpm=200\gev$ and
$\tanb=50$ ($\tanb=20$), then the $1\sigma$ errors (including systematic
uncertainties) on $\tanb$ 
are $\Delta\tanb/\tanb=0.06$ ($\Delta\tanb/\tanb=0.2$), respectively,
for $\call=2000\fbi$ and $\rts=500\gev$. 

\begin{figure}[h!]
 \begin{minipage}{0.75\textwidth}
\begin{center}
\hspace*{-.8in}
\includegraphics[width=1.0\textwidth,angle=90]{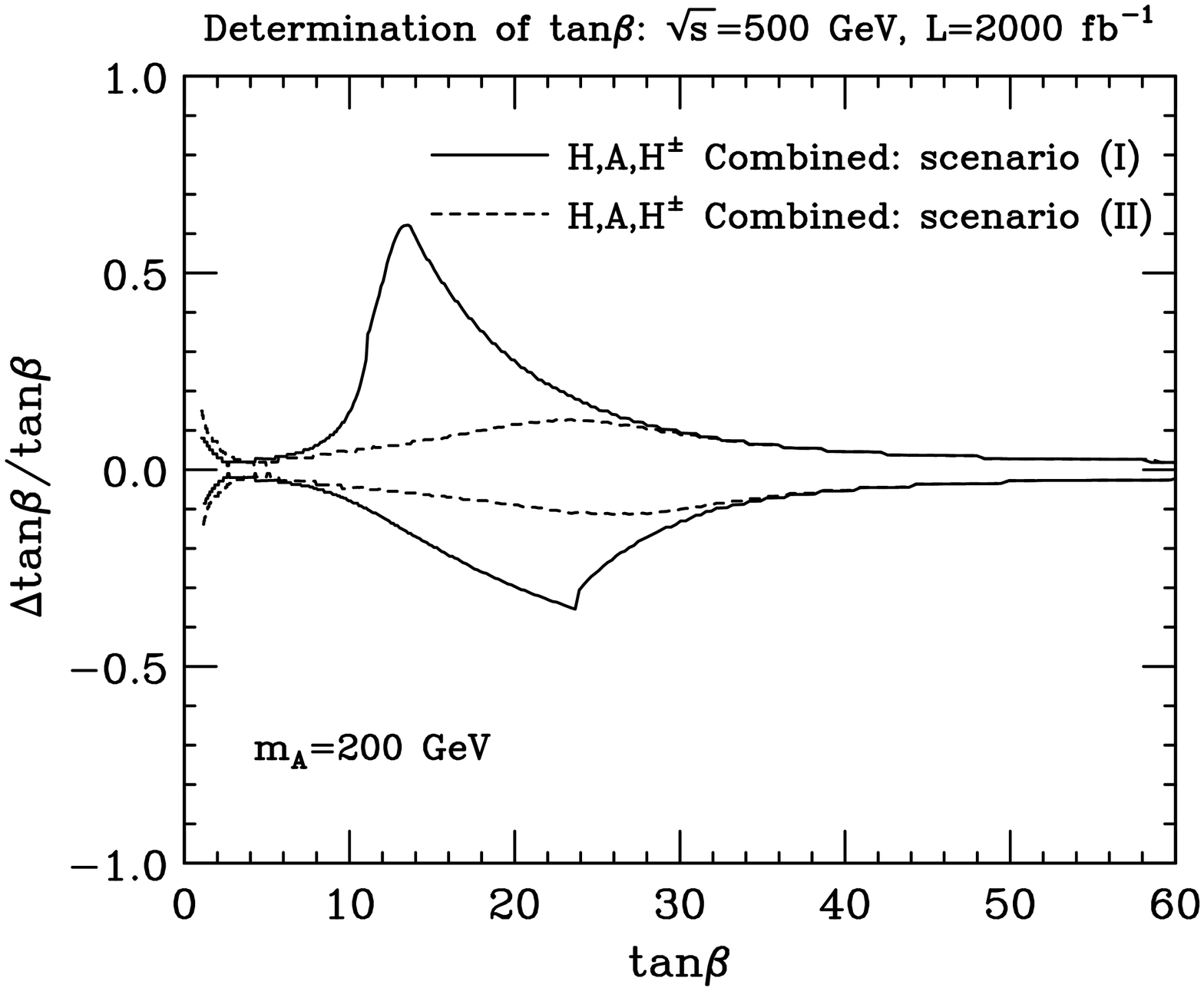}
\end{center}
\end{minipage}
\caption{\label{totalonly}
For the MSSM with $\mhpm\sim \mha = 200$~GeV, and 
assuming $\call=2000\fbi$ at $\rts=500\gev$,
we plot the $1\sigma$ statistical upper and lower bounds, $\Delta\tanb/\tanb$,
as a function of $\tanb$ based on combining (in quadrature)
the results shown in Figs.~\ref{hhhaonly} and \ref{hpmonly}.
Results are shown for the SUSY scenarios (I) and (II) described
in the text.}
\end{figure}

By combining in quadrature the $\tanb$ errors for the various individual
measurements, as given in Figs.~\ref{hhhaonly}
and \ref{hpmonly}, 
we obtain the net errors on $\tanb$ shown in Fig.~\ref{totalonly}.

\vspace*{-4mm}
\section{\boldmath Comparison to LHC determinations of $\tanb$}
\vspace*{-4mm}

In this section, we will compare
the LC results summarized in Figs.~\ref{hhhaonly}, \ref{hpmonly}
and \ref{totalonly} to the $\tanb$ 
accuracies that can be achieved at the LHC
based on $\hh,\ha,\hpm$ production and decay processes.  
First note that there is a wedge-shaped window of moderate $\tanb$
and $\mha\gsim 200\gev$ for which the $\ha$, $\hh$
and $\hpm$ are all unobservable (see, for example,
Refs.~\cite{LHCcms,LHCatlas,Gianotti:2002xx}). In this wedge,
the only Higgs boson that is detectable at the LHC is
the light SM-like  Higgs boson, $\hl$. Precision measurements of the
properties of the $\hl$ typically only provide weak sensitivity
to $\tanb$, and will not be considered here. The lower $\tanb$
bound of this moderate-$\tanb$ wedge is defined by the LEP-2
limits~\cite{lepwg}, which are at $\tanb\sim 3$ for $\mha\sim 200\gev$, falling
to $\tanb\sim 2.5$ for $\mha\gsim 250\gev$, assuming
the maximal mixing scenario [see SUSY scenario (I) defined earlier].
The upper $\tanb$ limit of the wedge is at $\tanb\sim 7$
for $\mha\sim 200\gev$ rising to $\tanb\sim 15$ at $\mha\sim 500\gev$. 
For either smaller or larger $\tanb$
values, the heavy MSSM Higgs bosons can be detected and 
their production rates and properties will provide sensitivity to $\tanb$.  

We will now summarize the 
results currently available regarding the determination of
$\tanb$ at the LHC using Higgs measurements (outside the wedge region)
assuming a luminosity of $\call=300\fbi$. The methods employed are
those proposed in \cite{Gunion:1996cn}.
The reactions that have been studied at the LHC are the following.
\ben
\item $\hh\to ZZ\to 4\ell$ \cite{LHCatlas}.

 The best accuracy that can be achieved
at low $\tanb$ is obtained from
the $\hh\to ZZ\to 4\ell$ rate. One finds $\Delta\tanb/\tanb=\pm 0.1$ 
at $\tanb=1$ rising to $>\pm 0.3$ by $\tanb=1.5$ for the sample
choice of $\mhh=300\gev$. For $\mhh<2m_Z$, $\tanb$
cannot be measured via this process. In the MSSM maximal mixing
scenario, such low values of $\tanb$
are unlikely in light of LEP-2 results.

\item $gg\to \hh+gg\to \ha\to \tau^+\tau^-,\mu^+\mu^-$ and
$gg\to b\anti b \hh+b\anti b \ha\to \bb\tau^+\tau^-,\bb\mu^+\mu^-$
\cite{LHCatlas}.

At high $\tanb$ and taking $\mha=150\gev$, Fig.~19-86 of  \cite{LHCatlas}
shows that 
the $gg\to \hh\to\tau^+\tau^-$, $gg\to \ha\to\tau^+\tau^-$, and
$gg\to\bb\ha+\bb\hh\to \bb\tau^+\tau^-$ rates can, in combination,
be used to determine
$\tanb$ with an accuracy of $\pm 0.15$ at $\tanb=5$, improving
to $\pm 0.06$ at $\tanb=40$. The corresponding rates with 
$\hh,\ha\to\mu^+\mu^-$ yield a somewhat better determination
at higher $\tanb$: $\pm 0.12$ at $\tanb=10$
and $\pm 0.05$ at $\tanb=40$. 

Interpolating, using Figs.~19-86 and 19-87 from \cite{LHCatlas},
we estimate that at $\mha\sim 200\gev$ (our choice for this study)
the error on $\tanb$ based on these rates 
would be smaller than $\pm 0.1$ for $\tanb\gsim 13$,
asymptoting to $\pm 0.05$ at large $\tanb$.

It is important to note that the $\tanb$ sensitivity 
for $\tanb<20-30$ is largely due to the loop-induced $gg\to \ha$
and $gg\to\hh$ production processes. Thus,  
interpreting these fully inclusive rates in terms of $\tanb$ 
when $\tanb<20-30$ requires
significant knowledge of the particles, including SUSY particles,
that go into the loops responsible for the $gg\to\hh$ and $gg\to\ha$ 
couplings. 

The importance of including
the $gg\to \hh$, $gg\to\ha$ as well as the $gg\to \bb\hh+gg\to \bb\ha$
processes in order to obtain observable signals
for $\tanb$ values as low as 10 
in the $\mu^+\mu^-$ channels is apparent from \cite{Dawson:2002cs}.
For $\call=300\fbi$ and $\mha=200\gev$, they find that
the $\bb\mu^+\mu^-$ final states can only
be isolated for $\tanb>30$ whereas the inclusive $\mu^+\mu^-$ final
state from all production processes becomes detectable once $\tanb>10$.

\item $gg\to t\anti b \hm +\anti t b\hp$ with $\hpm\to \tau^{\pm}\nu$
\cite{Assamagan:2002ne}.

The $tb\hpm\to tb\tau\nu$ rate gives a 
fractional $\tan\beta$ uncertainty, $\Delta\tanb/\tanb$,
ranging from $\pm 0.074$ 
at $\tanb=20$ to $\pm 0.054$ at $\tanb=50$. This signal is somewhat
cleaner to interpret in terms of a $\tanb$
measurement than the inclusive signals for the $\hh$
and $\ha$ summarized above, since there are no uncertainties
related to SUSY loop contributions.
 
\een
Sensitivity
to $\tanb$ deriving from direct measurements of the decay widths
has not been studied by the LHC experiments.
One can expect excellent $\tanb$ accuracy at the higher $\tanb$ values
for which the $gg\to b\anti b \mu^+\mu^-$ signal for the $\hh$
and $\ha$ is detectable.

\begin{table}[b!]
\centering
\caption{\label{summarytable}A comparison of fractional errors, 
$\Delta\tanb/\tanb$, achievable for $\call=2000\fbi$
at the LC with those expected at the LHC for $\call=300\fbi$,
assuming $\mha=200\gev$ in the MSSM. LC results are given for both 
SUSY scenarios (I) and (II),
where Higgs boson decays to SUSY particles are disallowed,
respectively allowed. LHC results are estimated
by roughly combining the determinations of $\tanb$ based on
$\hh,\ha$ production from \cite{LHCatlas}
with those using $\hpm$ production from \cite{Assamagan:2002ne},
both of which assume the standard MSSM
maximal mixing scenario. All entries are approximate.}
\vskip6pt
\begin{tabular}{|c|c|c|c|}
\hline
 $\tanb$ range &    LHC           &       LC (case I)  &      LC (case II) \cr 
\hline
 1       &  0.12       &  0.15        &   0.1         \cr
 1.5--5  &  very large &  0.03--0.05  &   0.03--0.05  \cr
 10      &  0.12       &  0.1         &   0.05        \cr
 13--30   &  0.05       &  0.6--0.1    &   0.05--0.1   \cr
 40--60   &  0.05--0.03 &  0.05--0.025 &   0.05--0.025 \cr
\hline 
\end{tabular}
\end{table}

Let us now compare these LHC results to the LC errors for $\tanb$,
assuming $\mha=200\gev$.
First, consider $\tanb\leq 10$. 
As summarized above, the LHC error on $\tanb$ is $\pm 0.12$
at $\tanb\sim 10$ and at $\tanb\sim 1$, and 
the error becomes very large for $1.5\lsim \tanb\lsim 5$.
Meanwhile, the LC error from Fig.~\ref{totalonly} 
ranges from roughly $\pm0.03$ to $\pm 0.05$ for $2\lsim\tanb\lsim 5$ rising to
about $\pm 0.1$ at $\tanb\sim 10$ [in the less favorable
SUSY scenario (I)]. Therefore, for $\tanb\lsim 10$ 
the LC provides the best determination of $\tanb$ using Higgs observables
related to their Yukawa couplings.
(In the MSSM context, 
other non-Higgs LHC measurements would allow a good $\tanb$ determination
at low to moderate $\tanb$ based on other kinds of couplings.)  
In the middle range of $\tanb$ (roughly $13<\tanb<30$ at $\mha\sim 200\gev$),
the heavy Higgs determination of
$\tanb$ at the LHC might be superior to that obtained at the LC.
This depends upon the SUSY scenario: if the heavy Higgs bosons can decay
to SUSY particles, the LC will give $\tanb$ errors that
are quite similar to those obtained at the LHC;
if the heavy Higgs bosons do not have substantial SUSY decays,
then the expected LC $\tanb$ errors are substantially larger than those
predicted for the LHC.
At large $\tanb$, the LC measurement of the heavy Higgs couplings
and the resulting $\tanb$ determination at the LC is numerically only
slightly more accurate than that obtained at the LHC.
For example, both are of order $\pm 0.05$ at $\tanb=40$.  
These comparisons are summarized in Table~\ref{summarytable}.
%If a direct measurement of the $\hh,\ha$ average width
%in the $b\anti b \mu^+\mu^-$ channel is included
%in the LHC analysis in the future, 
%it is possible that the net LHC $\tanb$ error would
%be somewhat smaller than the LC error for $\tanb\gsim 40$.However, we must
It is possible that the net LHC $\tanb$ error would
be somewhat smaller than the LC error for $\tanb\gsim 40$
if both ATLAS and CMS can each accumulate
$\call=300\fbi$ of luminosity; combining the two data sets would
presumably roughly double the statistics and decrease errors
by a factor of order $1/\sqrt 2$.  In any case, 
a very small error on $\tanb$ will be achievable for all $\tanb$
by combining the results from the LC with those from the 
LHC.~\footnote{However, we emphasize
that the above LHC versus LC comparisons have
been made based only on $\hh,\ha,\hpm$ processes and
for the particular choice of $\mha=200\gev$ in the MSSM, 
assuming $\rts=500\gev$ for the LC.
It is very possible that the relative accuracies as a function
of $\tanb$ could be quite different if the LC energy
is such that $\mha>\rts/2$, since then $\hh\ha$ and $\hp\hm$ 
pair production would not be possible.}

\vspace*{-4mm}
\section{Conclusions}
\vspace*{-2mm}

A high-luminosity linear collider will provide 
a precise determination of the value of $\tanb$
throughout much of the large range of possible interest, $1<\tanb<60$.
In this paper, we have studied the sensitivity
to $\tanb$ that will result from measurements of
heavy Higgs boson production processes,
branching fractions and decay widths. These are all directly
determined by the ratio of vacuum expectation values that defines
$\tanb$, and each can be very accurately measured at an LC
over a substantial range of relevant $\tanb$ values.
In particular, there are several
Higgs boson observables which are likely to provide the most precise
measurement of $\tanb$ when $\tanb$ is very large. 
%
%The highly-precise measurements of Higgs boson production
%processes and properties possible at 
%a high-luminosity linear collider will allow
%a uniquely precise determination of the value of $\tanb$. 
In the context of the MSSM, there is a particularly large variety
of complementary methods that will allow an accurate determination of $\tanb$ 
when $\mha\lsim \rts/2$ so that $\epem\to\hh\ha$
pair production is kinematically allowed.  
Using the sample case of $\mha=200\gev$
(in the MSSM context) and a LC with $\rts=500\gev$,
we have demonstrated the complementarity of employing: 
\begin{itemize}
\itemsep=0in
\item[a)] the $\bb\ha,\ \bb\hh \to \bbbb$ rate; 
\item[b)] the $\hh\ha\to \bbbb$ rate; 
\item[c)] a measurement of the average $\hh,\ha$ total width in $\hh\ha$ production;
\item[d)] the $\hp\hm\to \tbtb$ rate; and 
\item[e)] the total $\hpm$ width measured in $\hp\hm\to\tbtb$ production.
\end{itemize}
By combining the $\tanb$ errors from all these processes in quadrature,
we obtain the net errors on $\tanb$ shown in Fig.~\ref{totalonly}
by the lines [solid for SUSY scenario (I) and dashed for SUSY scenario (II)],
assuming a multi-year integrated luminosity of $\call=2000\fbi$.
We see that, independent of the scenario, 
the Higgs sector will provide an excellent determination of $\tanb$
at small and large $\tanb$ values,
leading to an error on $\tanb$ of $10\%$ or better. If SUSY decays
of the $\hh,\ha,\hpm$ are significant [SUSY scenario (II)], the $\tanb$ error 
will be smaller than $13\%$ even in the more difficult moderate $\tanb$ range.
However, if SUSY decays are not significant
[SUSY scenario (I)] there is a limited range of moderate $\tanb$
for which the error on $\tanb$ would be large, reaching about $50\%$.

In the preceding section, we considered how these $\tanb$
errors from the LC compared to $\tanb$ errors determined
at the LHC based only on measurements involving $\hh,\ha,\hpm$
production and decay.  The broad conclusions were: (i)
for low $\tanb$ ($\lsim 10$) the errors on $\tanb$ from LHC
Higgs measurements would be much larger than those attainable at the LC;
(ii) for high $\tanb$ ($\gsim 30$) the LHC and LC $\tanb$
errors were both small and quite comparable in magnitude; and (iii)
in the moderate-$\tanb$ range ($13\lsim \tanb\lsim 30$)
the LHC errors on $\tanb$ would very possibly be smaller than the LC errors.
However, we also noted that in this latter region
some care in interpretation of the LHC results would be necessary due to the
need to include loop-induced $gg\to \hh$ and $gg\to \hh$ production processes
in order to obtain good sensitivity to $\tanb$; these
might be influenced by loops of SUSY particles and, possibly, other
undiscovered new physics. This LHC versus LC comparison should also
be viewed as highly preliminary since the LHC collaborations
have not yet studied all the relevant observables.  In particular,
they have not looked at the $\tanb$ determination using the directly
measured widths of the $\hh$ and $\ha$.
Regardless of the relative magnitude of the LHC versus LC
$\tanb$ errors, the clean LC environment will provide an important and
independent measurement that will complement any LHC determination of $\tanb$.
Different uncertainties will be
associated with  the determination of $\tanb$ at a
 hadron and an $e^+e^-$ collider because of the different backgrounds.
Further, the LHC and LC measurements of $\tanb$ will be highly complementary
in that the systematic errors involved will be very different.

Combining all the different LC measurements
as above does not fully account for the fact
that the ``effective'' $\tanb$ value being measured in each process
is only the same at tree-level. The $\tanb$ values
measured via the $\hh\to \bb$ Yukawa coupling, the $\ha\to\bb$ Yukawa
coupling and the $\hp\to t\anti b$ Yukawa coupling could all
be influenced differently by the MSSM one-loop corrections.   
For some choices of MSSM parameters, 
the impact of MSSM radiative corrections on
interpreting these measurements can be substantial \cite{Carena:1998gk}.
However, if the masses of the SUSY particles are known, so that
the important MSSM parameters entering these radiative
corrections (other than $\tanb$) are fairly well determined, then
a uniform convention for the definition of $\tanb$ can be adopted
and, in general, an excellent determination of $\tanb$ (with
accuracy similar to that obtained via our tree-level procedures)
will be possible using the linear collider observables considered here.
Even for special SUSY parameter choices such
that one of the Yukawa couplings happens to be significantly
suppressed, the observables a)-e) would provide an excellent
opportunity for pinning down all the Yukawa couplings 
and checking the consistency of the MSSM model.

Finally, it is important to note that 
the techniques considered here can also be employed
in the case of other Higgs sector models.
For example, in the general (non-SUSY) 2HDM, if the only non-SM-like 
Higgs boson with mass below $\rts$ is the $\ha$ \cite{Chankowski:2000an}, 
then a good determination
of $\tanb$ will be possible at high $\tanb$ from the 
$\bb\ha\to\bbbb$ production rate.
Similarly, in models with more than two Higgs doublet
and/or triplet representations, the
Yukawa couplings of the Higgs bosons, and, therefore, the analogues of
the 2HDM parameter $\tanb$, 
will probably be accurately determined through Higgs
production observables in $\epem$ collisions. 

\bigskip
\centerline{\bf Acknowledgments}
\smallskip
We wish to thank Ron Settles for discussions regarding
the expected detector resolution.
This work was supported in part by the U.S.~Department of Energy
under grants DE-FG02-95ER40896, DE-FG03-91ER40674, and W-31109-ENG-38,
and in part by
the Davis Institute for High Energy Physics, the Wisconsin Alumni
Research Foundation, and the Particle Physics 
and Astronomy Research Council, UK.

%\vspace*{-3mm}

\end{document}